\documentclass[pre,twocolumn,amssymb,superscriptaddress]{revtex4-2}
\usepackage{hyperref,amsmath,graphicx,subfigure}
\usepackage{geometry}
\newcommand{\arXiv}[1]{\href{http://www.arXiv.org/abs/#1}{arXiv:#1}}
\renewcommand{\doi}[2]{\href{http://dx.doi.org/#2}{#1}}

\begin{document}

\title{Random matrix theory of integrability-to-chaos transition}
\author{Ben Craps}
\affiliation{Theoretische Natuurkunde, Vrije Universiteit Brussel (VUB) and International Solvay Institutes, Brussels, Belgium}
\author{Marine De Clerck}
\affiliation{DAMTP, University of Cambridge, Cambridge, United Kingdom}
\author{Oleg Evnin}
\affiliation{High Energy Physics Research Unit, Faculty of Science, Chulalongkorn University, Bangkok, Thailand}
\affiliation{Theoretische Natuurkunde, Vrije Universiteit Brussel (VUB) and International Solvay Institutes, Brussels, Belgium}
\author{Maxim Pavlov}
\affiliation{Theoretische Natuurkunde, Vrije Universiteit Brussel (VUB) and International Solvay Institutes, Brussels, Belgium}

\begin{abstract}
The statistics of gaps between quantum energy levels is a hallmark criterion in quantum chaos and quantum integrability studies. The relevant distributions corresponding to exactly integrable vs.\ fully chaotic systems are universal and described by the Poisson vs.\ Wigner-Dyson curves. In the transitional regime between integrability and chaos, the distributions are much less universal and have not been understood quantitatively until now.
We point out that the relevant statistics that controls these distributions is that of the matrix elements of the nonintegrable perturbation Hamiltonian in the energy eigenbasis
of the unperturbed integrable system. With this insight, we formulate a simple random matrix ensemble that correctly reproduces the level spacing distributions
in a variety of test systems. For the distribution of matrix elements appearing in our construction, we furthermore discover surprising universal 
features: across a variety of physical systems with diverse degrees of freedom, these distributions are dominated by simple power laws.
\end{abstract}

\maketitle

\section{Introduction}\label{sec1}

The history of theoretical science revolves around identifying simple solvable models that describe natural phenomena.
More complex dynamical behaviors may then be analyzed in terms of how they deviate from these simple solvable models. 
In this way, the development of classical mechanics was inextricably linked
with the solution of the gravitational two-body problem. 
Likewise, quantum mechanics gained its shape side-by-side with the analytic solution for the
atomic spectrum of hydrogen.

In classical physics, solvability is understood in terms of the presence of a large number of conservation laws.
This picture first emerged as Liouville integrability, and developed later into the modern formats such as Lax integrability and beyond \cite{BBT}. In quantum physics, a similarly comprehensive
picture is lacking \cite{noint}, but recent decades witnessed rapid developments in the fields of quantum integrability and quantum chaos. 
One important question is which features of quantum systems connect to the integrable vs.\ chaotic dynamics of their classical counterparts.

A key ingredient of such considerations is the statistics of (appropriately rescaled or `unfolded') distances between neighboring energy levels of quantum systems \cite{DKPR}. 
The energy levels of typical integrable Hamiltonians are uncorrelated and the level spacings follow the Poisson distribution \cite{btint}
\begin{equation}
    P_P(s) = e^{-s}.
    \label{eq: Poisson distribution}
\end{equation}
In contrast, the level spacings of quantum systems with a chaotic classical limit are well-approximated \cite{BGS}  by {\it random matrix theory}. In the presence of time-reversal symmetry, this leads to the Gaussian Orthogonal Ensemble (GOE) whose level spacing statistics is governed by the Wigner-Dyson (WD) distribution accurately captured by the `Wigner surmise'
\begin{equation}
    P_{W}(s)=\frac{\pi s}2\, e^{-\pi s^2/4}.
    \label{eq: Wigner Dyson surmise}
\end{equation}
There is a characteristic `level repulsion' (vanishing probability density for vanishing gap size). Real-life manifestations of the WD universality are prominent in the physics of heavy nuclei \cite{nuclei} and molecules \cite{atomic}. The subject is experiencing an active revival at the moment \cite{AKMP,SRT} through applications of these ideas to open quantum systems, where the analogs of distributions \eqref{eq: Poisson distribution} and \eqref{eq: Wigner Dyson surmise}  are known \cite{GHS}, including direct connections to experiments \cite{openexp}.

The above descriptions are universal in the sense that systems that are classically integrable/chaotic develop energy levels upon quantization that follow these distributions. By contrast, no similarly simple universal statements can be made for the level spacings of systems transitioning from integrability to chaos when
an integrable quantum Hamiltonian is perturbed by a chaotic deformation. Different shapes of {\it intermediate} level spacing distributions can be seen in this regime depending on the concrete system one studies.
It is a natural expectation that a limited set of statistical properties of the physical Hamiltonian matrices control the shape of these distributions, but the literature
does not supply any explicit prescriptions in this regard, despite the considerable attention that has been paid to such {\it crossover} regimes interpolating between integrability and chaos. Our purpose in this article is to specify a class of random matrix ensembles that accurately reproduce the level spacing statistics during integrability-to-chaos transitions across a variety of physical Hamiltonians, and to test its performance. 

\section{Approaches to the crossover level statistics}

Studies of level spacing statistics away from pure integrability and full chaos have been conducted from a variety of perspectives. Early efforts focused on nuclear and atomic spectra that did not match the two universal descriptions outlined above \cite{brody,RP}. The emphasis was on developing phenomenological curves to be fitted to the actual data, rather than having a detailed theory. One such distribution was introduced by Brody \cite{brody} as
\begin{equation}
    P_{B,\beta}(s)
\sim s^{\beta}
e^{-a s^{1+\beta}}.
    \label{eq: Brody distribution}
\end{equation}
This curve both implements a `fractional level repulsion' at $s\rightarrow 0$ and contains an exponentially decaying factor at large $s$ so as to reduce to the Poisson distribution \eqref{eq: Poisson distribution} and the Wigner surmise \eqref{eq: Wigner Dyson surmise} at $\beta = 0$ and $\beta=1$, respectively.  Brody's distribution provides a good fit for the intermediate statistics in some models \cite{brodyEx,banded1}, in particular, in the regime away from the semiclassical limit in certain billiards \cite{BR3,BR5}, while it fails for some spin chains \cite{gamma}, and for the simple systems we consider below. The parameter $\beta$ has no direct physical interpretation, and  is typically used for fitting to obtain an estimate of how far the system is away from integrability. 

The second benchmark we consider is the Rosenzweig-Porter (RP) model \cite{RP}, defined as an ensemble of random $D \times D$ matrices of the form
\begin{equation}
    \mathcal{H}_{RP} = \mathcal{D}_{RP} + \gamma_{RP}\, \mathcal{M}_{GOE}.
    \label{eq: def RP}
\end{equation}
Here, $\mathcal{D}_{RP}$ is a diagonal matrix with independent and identically distributed (i.i.d.) entries governed by a normal distribution $\mathcal{N}(0,1)$, and  $\mathcal{M}_{GOE}$ is drawn from the GOE ensemble (all the entries are independent Gaussian variables). The crossover parameter $\gamma_{RP}$ can be used to dial the intensity of the GOE perturbation, inducing a transition between level spacing distributions \eqref{eq: Poisson distribution} and \eqref{eq: Wigner Dyson surmise}. This model was originally proposed to describe certain atomic spectra, and has regained interest in the last decade due to the appearance of fractal states during the transition \cite{RPfractal}. In relation to level spacing statistics, the RP model has mostly been used for studies of qualitative features of the crossover regime, and is known to fail quantitatively in concrete examples \cite{RPfail}, which will also be seen immediately in our numerics below.

Besides fitting formulas like the Brody distribution and basic random matrix ensembles like the RP model, there have also been attempts at constructing an analytic theory of the transition from Poisson to WD statistics in physical systems. One approach originated from the semiclassical perspective of Percival's \cite{Percival},
which purports that the spectrum of a Hamiltonian interpolating between integrability and chaos separates into `integrable' and `chaotic' eigenstates, see also \cite{WnW}. This separation mimics the splitting of classical phase spaces into regions of regular and chaotic motion. Berry and Robnik \cite{BR1} used this idea to derive the intermediate statistics for a spectrum composed of statistically independent sequences of levels, some of them following \eqref{eq: Poisson distribution} and others \eqref{eq: Wigner Dyson surmise}. Their proposed distribution was found to successfully describe the crossover statistics in various models deep in the semiclassical regime \cite{BR2,BR3}. However, away from the semiclassical regime or in models with no classical limit, the Berry-Robnik distribution appears ineffective and, in particular, fails at small $s$, missing the characteristic level repulsion \cite{BR3}. Some further approaches to the crossover level spacing statistics can be seen in
 \cite{RPfail,crossoverCollection}, but their degree of success is model-specific.
We shall now proceed with constructing a simple random matrix ensemble that provides for systematic control of the crossover behaviors.

\section{A random matrix ensemble for integrability-to-chaos transition}\label{sec2}

We aim to understand the level spacing statistics for physical models with $D$-dimensional Hilbert spaces and Hamiltonians
of the form
\begin{equation}
   H =  H_0 + \gamma H_1,
   \label{eq: H}
\end{equation}
where $H_0$ is an integrable Hamiltonian and $H_1$ is a chaotic perturbation (with WD statistics of its level spacings), while $\gamma$ is a tunable parameter. The level spacing distribution of \eqref{eq: H} transitions from Poisson at $\gamma = 0$ to WD at large $\gamma$. 

The main result we report here is that the level spacing distribution of \eqref{eq: H} is controlled, at large $D$, by the statistics of the matrix elements of $H_1$ in the eigenbasis of $H_0$. In other words, the precise positions and values of the entries of $H_1$ in the eigenbasis of $H_0$ are irrelevant for our purposes, and the distribution of the total, unordered sample of values within the matrix of $H_1$ is what one needs to know. To express this statement mathematically, we formulate the following random matrix ensemble: 
\begin{equation}
    \mathcal{H} = \mathcal{D}+\gamma'\mathcal{M},
    \label{eq: RM ensemble}
\end{equation}
where $\mathcal{D}$ and $\mathcal{M}$ are a diagonal matrix and a zero-diagonal symmetric matrix respectively, with i.i.d.\ nonzero entries drawn from two distinct probability distributions $P_{\mathcal{D}}$ and $P_{\mathcal{M}}$ extracted from $H_0$ and $H_1$, as we shall now describe. With $|n\rangle$ denoting the energy eigenstates of $H_0$,
\begin{equation}
H_0|n\rangle=E_n|n\rangle,
\end{equation}
these distributions are obtained by appropriate smoothing from the following two large samples:
First, for $P_{\mathcal{D}}$, we take the total sample of $H_0$ eigenenergies $E_n$. Second, $P_{\mathcal{M}}$ is extracted from the total sample of offdiagonal matrix elements $H_{1,nm} = \langle n|H_1|m\rangle$ with all $n<m$. (We discard the diagonal matrix elements of $\mathcal{M}$ since the crossover typically happens for small values of $\gamma'$ of order $1/D$, see the discussion for the RP model in \cite{LS,RPreplica}, and in this regime, their contribution to the diagonal is small relative to the contribution of $\mathcal{D}$.) We outline the protocol for extracting $P_{\mathcal{D}}$ and $P_{\mathcal{M}}$ in more detail in the appendix.

Our claim is that ensemble (\ref{eq: RM ensemble}) correctly reproduces the level spacing distribution of the physical model (\ref{eq: H}). Before proceeding with verifying this claim, we need to dwell on two important technical points. First, the definition of level spacing distributions involves the important step of {\it spectral unfolding} \cite{BGS,unfolding}, which makes sure that the average level spacing (over ranges much smaller than the full extent of the spectrum) does not depend on the position within the spectrum. This allows for meaningfully joining together the different energy ranges with different average level densities, obtaining a distribution that controls the level fluctuations, and comparing such distributions between different systems. Second, we need to specify how to identify the deformation parameters $\gamma$ and $\gamma'$ appearing in (\ref{eq: H}) and (\ref{eq: RM ensemble}), so as to make sure that similar stages across the transition are compared to each other. We will employ the convenient technique of tracking the average $r$-ratios \cite{rratio1,rratio2} to this end. Both points are addressed in the following two sections.

Before we proceed, it is useful to note that, in the absence of $\mathcal{D}$ in \eqref{eq: RM ensemble}, the shape of the distribution $P_{\mathcal{M}}$ would not matter at all, which is indeed
one of the key universal properties leading to the WD statistics. When $\mathcal{D}$ is present, however, the level spacing distributions become sensitive to the shape
of $P_{\mathcal{M}}$ in the crossover regime (that is, when $\gamma'$ is such that the two terms in \eqref{eq: RM ensemble} are equally significant). The crossover
regime is then described by a large family of distinct universality classes defined by \eqref{eq: RM ensemble}, while further details (such as possible correlations
between the matrix entries  of $\mathcal{M}$) become irrelevant as seen in our numerical experiments below.

\section{Unfolded level spacings}

To compute the level spacing distribution, we first need to unfold the spectrum so as to make the level density uniform over different `mesoscopic' intervals (many levels within each interval, but still much fewer than the total number of levels). To implement this idea, we resort to the following simple procedure going back to \cite{quantres}: we first remove $\lfloor D^{1/2} \rceil$ eigenvalues located near the edges of the spectrum, leaving $\tilde{D}=D-2\lfloor D^{1/2} \rceil$ eigenvalues to work with. (The spectral edge often shows nongeneric features and a slower progress toward chaos \cite{lenzhaake}.) We then introduce a `mesoscopic' scale $\Delta = \lfloor \tilde D^{1/2} \rceil$, and define the unfolded spacings $s_n$ as
\begin{equation}
    s_n^{(\mathrm{raw})}
= \frac{E_{n+1} - E_n}{E_{n+\Delta} - E_{n-\Delta}}, \quad s_n = \frac{s_n^{(\mathrm{raw})}}{\bar{s}^{(\mathrm{raw})}} \, ,
\label{eq: unfolding}
\end{equation}
where the normalization by the average
${\bar{s}^{(\mathrm{raw})}} \equiv \sum_{n=\Delta+1}^{\tilde{D}-\Delta} s_n^{(\mathrm{raw})}/(\tilde{D} - 2\Delta)$
ensures that the mean level density is 1.
We have compared this straightforward method against the more commonly used polynomial unfolding \cite{polunfolding}, and found little difference.
In the analysis below, the spectrum of each Hamiltonian (physical and drawn from one of the random matrix ensembles) will be unfolded according to \eqref{eq: unfolding} and we will be interested in the statistics of the variables $s_n$. 

\section{The $r$-ratio}

Our main claim is that the level spacing statistics match between (\ref{eq: H}) and (\ref{eq: RM ensemble}) under appropriate identification of the perturbation parameters $\gamma$ and $\gamma'$. To provide this identification, we need empirical input in the format of a single number (quantifying how much overall change in the spectrum is induced by the given values of $\gamma$ and $\gamma'$). This extra input then allows us to pin down the entire level spacing distribution curve.

To quantify the effect of perturbations in (\ref{eq: H}) and (\ref{eq: RM ensemble}), we inspect the average $r$-ratio that has been used extensively in quantum chaos studies. To define it, for a given spectrum $E_n$, indexed in ascending order, after removing the edges of the spectrum, consider
\begin{equation}
\label{eq: r-ratio definition}
r_n
\equiv \frac{\min(\delta_n, \delta_{n-1})}{\max(\delta_n, \delta_{n-1})} \, ,
\end{equation}
with $\delta_n\equiv E_{n+1}-E_n$. The distribution of the $r$-ratios for dynamical models was first proposed in \cite{rratio1} as an attractive way to characterize the chaotic properties of a Hamiltonian that does not require unfolding, and we will comment more on this in the appendix. What is important for us here is the average value of $r_n$ over the spectrum, denoted as $r$, which is known for Poisson-distributed eigenvalues $ r_P = 2 \log 2 - 1 \approx 0.386$
\cite{rratio1} and in the GOE ensemble $r_{WD} = 4- 2 \sqrt{3}\approx 0.5359 $
 \cite{rratio2}. As the crossover is taking place, we will parametrize $\gamma$ and $\gamma'$ in (\ref{eq: H}) and (\ref{eq: RM ensemble}) through the average $r$-ratio they induce as $\gamma(r)$ and $\gamma'(r)$. We will then be comparing the level spacing distributions at the same value of $r$.

\section{Validation for spin 1/2 chains}
\label{sec:test systems}

\begin{figure*}[t]
    \centering
   \includegraphics[scale=0.6,trim= 0.5cm 0.5cm 0.5cm 0.5cm,
  clip]{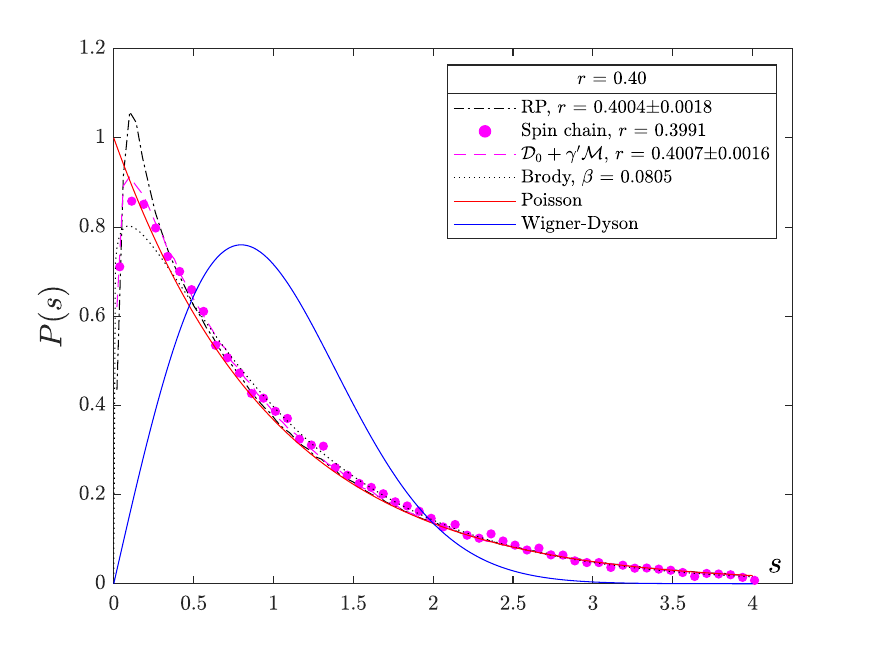} \includegraphics[scale=0.6,trim= 0.5cm 0.5cm 0.5cm 0.5cm,
  clip]{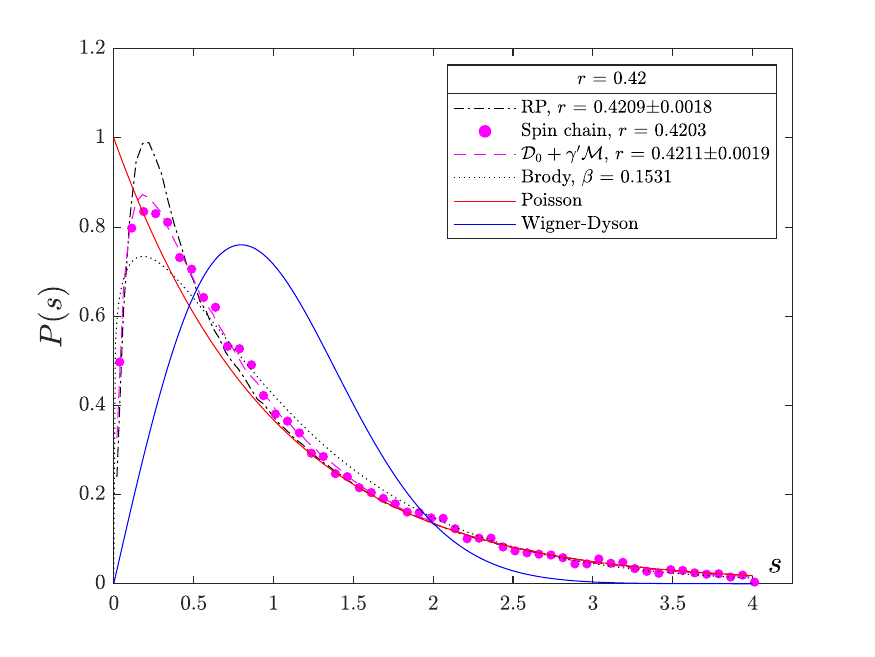}\\
  \includegraphics[scale=0.6,trim= 0.5cm 0.5cm 0.5cm 0.5cm,
  clip]{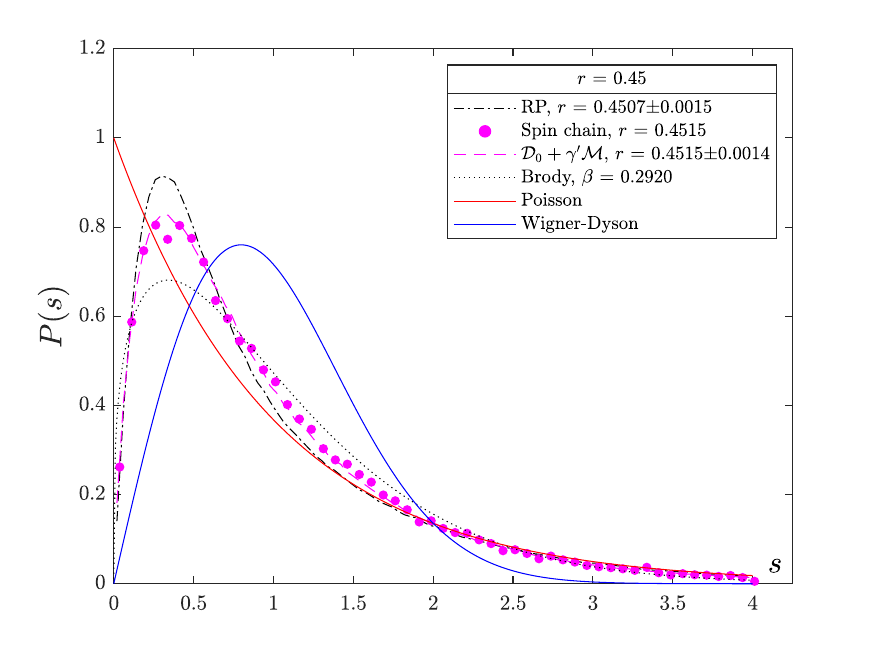} \includegraphics[scale=0.6,trim= 0.5cm 0.5cm 0.5cm 0.5cm,
  clip]{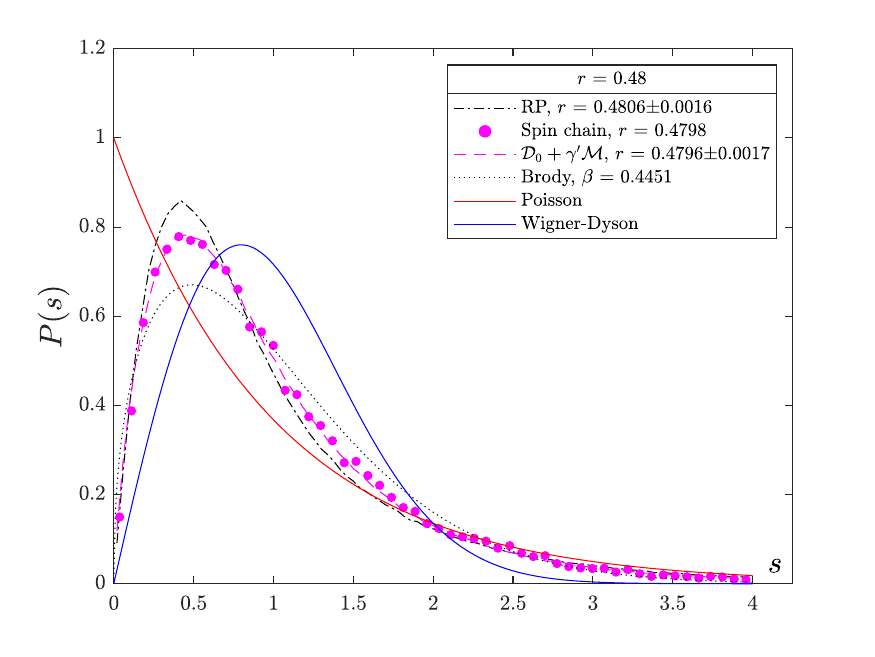}
    \caption{The transition in level spacing distribution from Poisson to WD at four different values of $r$ for the spin chain model (\ref{eq: H0 spin chain}-\ref{eq: H1 spin chain}) at $L = 16$ plotted as filled circles, compared with the joint distribution of spacings for 50 realizations of the random matrix ensemble (\ref{eq: RM ensemble})  in dashed magenta. The Hilbert space dimension is $D=32896$. As a matter of comparison, we also display the Poisson distribution \eqref{eq: Poisson distribution} in red, the Wigner surmise \eqref{eq: Wigner Dyson surmise} in blue, the level spacing statistics of the RP model, based on 50 realizations at $D= 32896$, at the same values of $r$ in dash-dotted black, and the Brody distribution as a dotted black curve, with $\beta$ found by fitting \eqref{eq: Brody distribution} to the physical level spacing distribution.}
    \label{fig: Ising histograms}
\end{figure*}
The first arena for verifying our construction is provided by the integrable transverse spin-$1/2$ Ising chain with open boundary conditions,
\begin{equation}
    H_0 = - \sum_{i=1}^{L-1} Z_iZ_{i+1} - \sum_{i=1}^LX_i,
    \label{eq: H0 spin chain}
\end{equation}
where $X_i, Y_i$ and $Z_i$ are the Pauli matrices acting on site $i$ (technically, restricting to the even parity sector with respect to reflection of the chain through its midpoint). 
We consider a chaotic perturbation of $H_0$ in the form of the XY-Heisenberg Hamiltonian with an external magnetic field:
\begin{align}
H_1 = &- \sum_{i=1}^{L-1} \left( J_{xx} X_iX_{i+1} +  J_{yy} Y_iY_{i+1}  \right)\nonumber 
    \label{eq: H1 spin chain}\\
   & - \sum_{i=1}^L(  J_x X_i+J_z Z_i ),
\end{align}
where the (spatially uniform) coefficients $J_{xx}$, $J_{yy}$, $J_{x}$ and $J_z$ are chosen randomly from  the interval $[-1/2,1/2]$. Spin chains are common in studies of integrability-to-chaos transitions in many-body systems \cite{crossover,santos,weakintegrability,crossover_SC}, especially in the context of many-body localization \cite{heisenberg,MBL,gamma}. Note that the Hamiltonian (\ref{eq: H1 spin chain}) is chaotic, so that the total Hamiltonian is of the form stated under (\ref{eq: H}).

We analyze the level spacing statistics for the sum of \eqref{eq: H0 spin chain} and a single random realization of \eqref{eq: H1 spin chain} across the transition from Poisson to WD. We focus on values of $\gamma$ and $\gamma'$ selected to produce $r = \{0.4, 0.42, 0.45,0.48\}$. After constructing $P_{\mathcal{D}}$ and $P_{\mathcal{M}}$ according to our general protocol, we sample random matrices from the ensemble (\ref{eq: RM ensemble}). For each of the values of $r$ listed above, we tune $\gamma'$ such that the average of $r$ over a large set of realizations of \eqref{eq: RM ensemble} at fixed $\gamma'$ matches that value.
Fig.~\ref{fig: Ising histograms} shows a comparison between the level spacing statistics of the physical system (\ref{eq: H0 spin chain}-\ref{eq: H1 spin chain}) and the joint distribution of $50$ realizations of the random matrix ensemble (\ref{eq: RM ensemble}), at these values of $r$. We contrast our approach with the performance of the RP model (also 50 realizations), where the free parameter of the model is fixed following the same procedure based on the value of $r$, and the Brody distribution, where the parameter $\beta$ is found via the best fit to the physical histogram. We have made public \cite{Zenodo} all the codes for this and other computations we are performing in this article. Our model shows excellent agreement with the empirical spectra, and clearly outperforms the models discussed in the past literature. 

\section{Validation for quantum resonant systems}

\begin{figure*}[t]
    \centering
    \includegraphics[scale=0.6,trim= 0.5cm 0.5cm 0.5cm 0.5cm,
  clip]{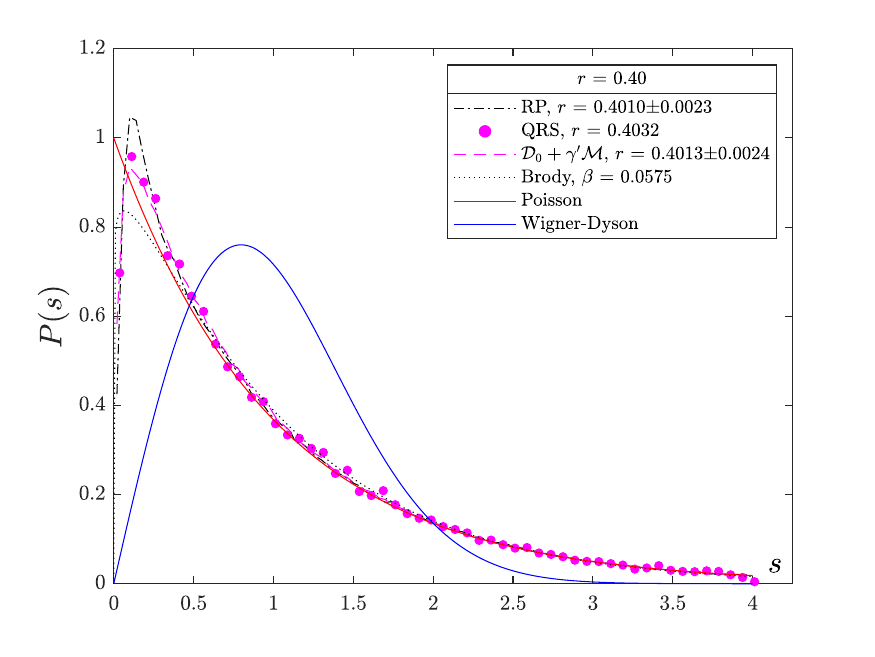} \includegraphics[scale=0.6,trim= 0.5cm 0.5cm 0.5cm 0.5cm,
  clip]{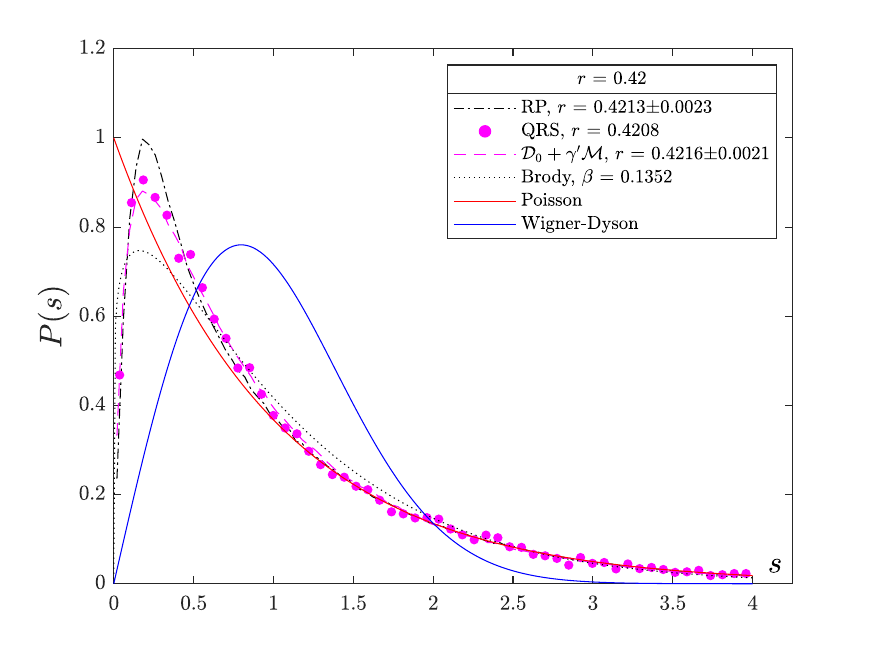}\\
  \includegraphics[scale=0.6,trim= 0.5cm 0.5cm 0.5cm 0.5cm,
  clip]{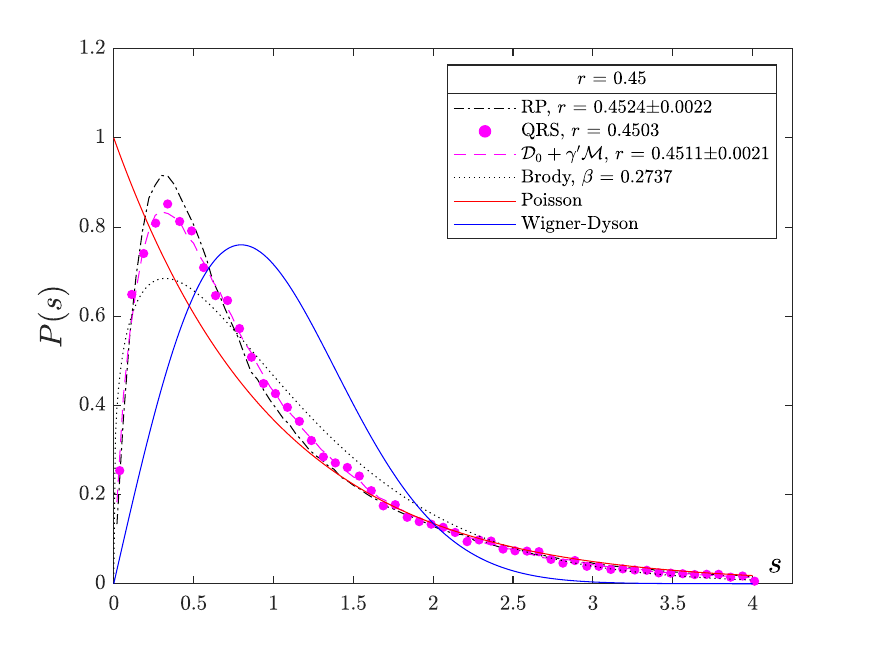} \includegraphics[scale=0.6,trim= 0.5cm 0.5cm 0.5cm 0.5cm,
  clip]{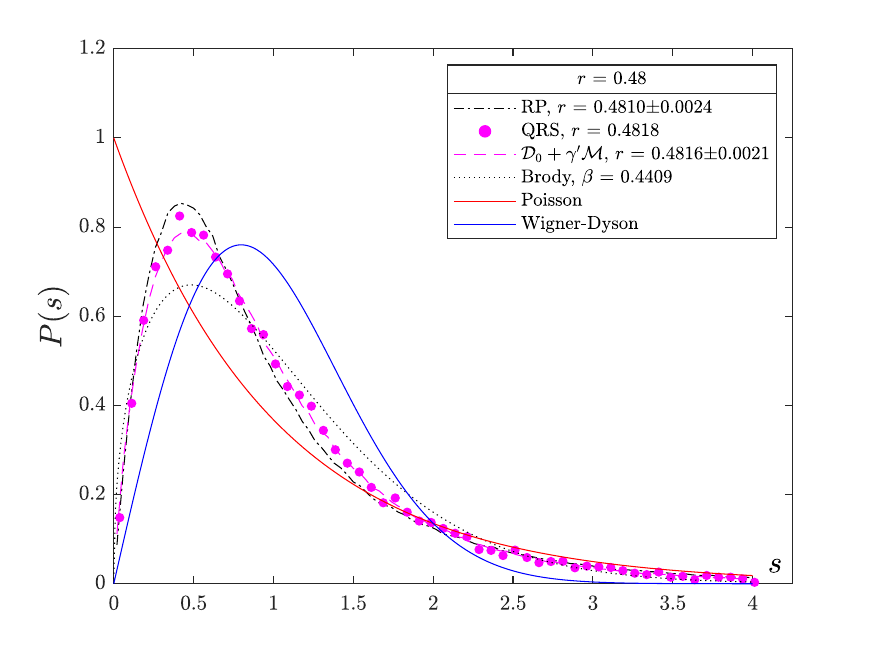}
    \caption{Similar to Fig.~\ref{fig: Ising histograms}, but now for the perturbed integrable QRS model \eqref{eq: truncated szego C} at $(N,M)=(37,37)$ with $D=21637$.}
    \label{fig: QRS histograms}
\end{figure*}
In order to test our ideas on another class of systems, we employ bosonic many-body models \cite{quantres} defined by the Hamiltonians
\begin{equation}
H=\frac{1}2\,\sum_{n,m,k,l\ge 0}^{n+m=k+l} \hspace{-2mm} C_{nmkl} a^\dagger_na^\dagger_m a_k a_l.
\label{eq: HQRS}
\end{equation}
The creation-annihilation operators for the modes labeled by $n\ge 0$ satisfy $[a_n,a^\dagger_m ] = \delta_{nm}$, while the interaction coefficients $C_{nmkl}$
are determined by the underlying physics. Such Hamiltonians may be referred to as {\it quantum resonant systems} (QRS), due to the resonance condition $n+m=k+l$ in \eqref{eq: HQRS}, and they emerge as weak coupling approximations 
to many physical systems with highly resonant normal mode spectra \cite{resrev}.
These Hamiltonians possess two conservation laws, 
$N=\sum_k a^\dagger_k a_k$ and $M=\sum_k k \, a^\dagger_k a_k$,
and as a result, the Fock space spanned by states 
$|\eta_0,\eta_1,\cdots \rangle$ with occupation numbers $\eta_k$, splits into finite-dimensional blocks indexed by $(N,M)$, where the Hamiltonian can be diagonalized as a finite-sized matrix \cite{quantres}. One thus has a bosonic system, with a well-defined classical limit and clear notions of integrability and chaos in this limit, that at the same time
behaves for practical purposes as if it had a finite number of independent quantum states. 

At  generic values of the couplings $C_{nmkl}$, the Hamiltonians \eqref{eq: HQRS} are chaotic and their level spacings follow the WD statistics \cite{quantres}. In contrast, rich quantum and classical integrable structures emerge for special choices of $C_{nkml}$ \cite{solvable,qperiod1,qperiod2,GGq}. As an integrable representative for $H_0$, we shall focus here on the QRS with interaction coefficients \cite{cascade}
\begin{equation}
    C^{(0)}_{nmkl} = 
    \begin{cases}
   0& \text{if none of the indices are 0}, \\
    1              & \text{otherwise}.
    \end{cases}
    \label{eq: truncated szego C}
    \end{equation}
This model exhibits Poissonian level spacing statistics \cite{complexity} and its classical limit is Lax-integrable \cite{cascade}. 
We perturb $H_0$ with a single random QRS realization, where the coefficients $C_{nmkl}$ in \eqref{eq: HQRS} are drawn from a uniform distribution over the interval $[0,1]$, while respecting the symmetries $C_{nmkl} =C_{mnkl} =C_{klnm} $. This choice defines $H_1$. 

With these definitions in hand, we repeat the numerical experiment of the previous section for QRS and display the results for the four values of $r$ in Fig.~\ref{fig: QRS histograms}. Here again, we find that the proposed random matrix ensemble captures the level spacing statistics of the physical Hamiltonian very well across the entire transition, and outperforms the previously considered models. 

\section{Statistics of the offdiagonal matrix entries of perturbations}\label{sec: offdiags}

\begin{figure*}[t]
    \centering
    \includegraphics[scale=0.6]{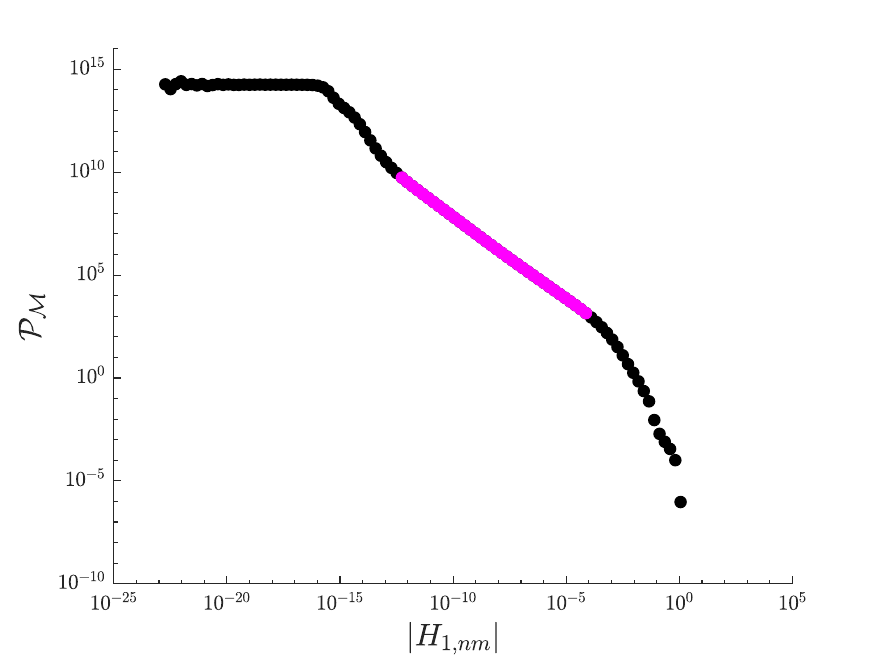}\includegraphics[scale=0.6]{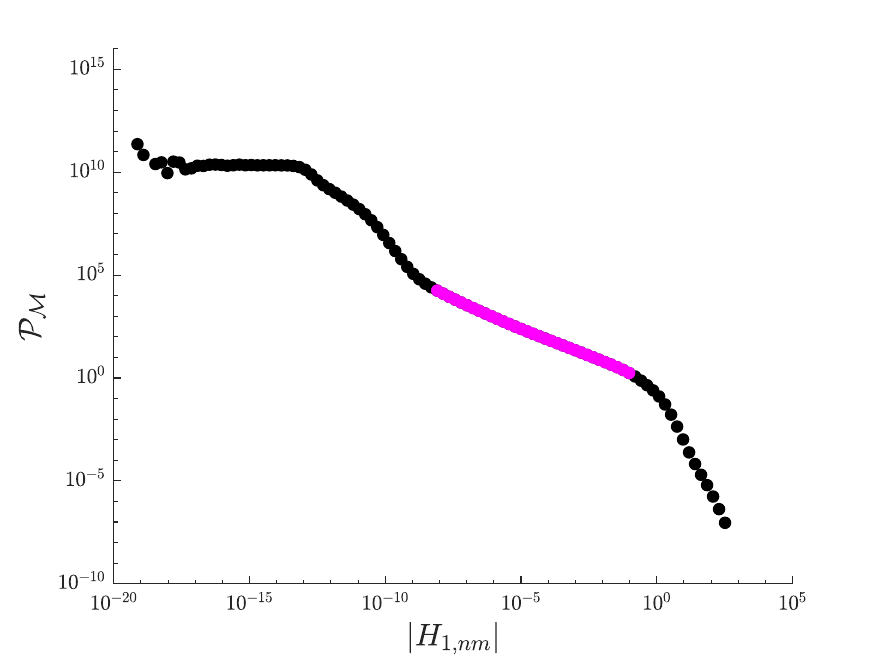}
    \caption{The probability density on a log-log scale for the magnitude of offdiagonal elements of the random perturbations $H_1$ in the eigenbasis of $H_0$ for \textbf{(left)} the spin chain model (\ref{eq: H0 spin chain}-\ref{eq: H1 spin chain}) at $L=16$ and \textbf{(right)} the perturbed integrable QRS model \eqref{eq: truncated szego C} at $(N,M)=(37,37)$. These distributions were used as $P_{\mathcal{M}}$ in our random matrix model approach to the crossover statistics. The log-log histograms reveal a power law regime (magenta) that extends over almost 10 orders of magnitude in both cases. The shape of the distributions outside of this regime plays no important role for the crossover statistics.}
    \label{fig: powerlaw distributions}
\end{figure*}
The distribution $P_{\mathcal{M}}$ of the offdiagonal elements of the random perturbation $H_1$ is a key ingredient in the random matrix ensemble \eqref{eq: RM ensemble}
we have formulated here. While the connection between this distribution and the level spacing statistics is valuable in itself, in our explorations of various test systems, we have discovered remarkable universal features of $P_{\mathcal{M}}$ that contribute to strengthening our construction further.

When examining the distribution $P_{\mathcal{M}}$ for spin chains and QRS models, we have observed the ubiquitous presence of power laws, as seen in Fig.~\ref{fig: powerlaw distributions}: 
each plotted distribution is dominated by a power law region that extends over many orders of magnitude, manifesting itself as a linear segment on the log-log histograms, while
deviations from this behavior only occur at very large and very small values. An instructive fitting formula for the curves in Fig.~\ref{fig: powerlaw distributions} is
\begin{equation}
P_{\mathcal{M}}(x)\sim \frac{e^{-Cx^2}}{x^p}.
\label{powerlw}
\end{equation}
The model-dependent power $p$, obtained from the slopes of the linear segments, lies in the interval $[0,1]$.
According to our observations, the precise form of the cutoff  ($e^{-Cx^2}$ in the above formula) and the deviations of the empirical distribution
from this fitting formula at very large and very small values have little effect on the corresponding level spacing distributions extracted from the ensemble \eqref{eq: RM ensemble}. Heuristically, the large entries deviating from the distribution (\ref{powerlw}) are too few, and the small entries are too small to have an effect on the level spacing statistics, which is dominated instead by the power-law region.
Similar power laws with $p>1$ have been prominently discussed in relation to L\'evy \cite{levy1,levy2,levy3} and L\'evy-RP \cite{levyRP} heavy-tailed random matrices, without connections to the statistics of matrix elements of physical Hamiltonians that we report here for $p<1$. Examples of other power laws related to quantum Hamiltonian dynamics can be seen in \cite{pwr1,pwr2,pwr3}.

To further support the universality of our observations, we have explored the distributions $P_{\mathcal{M}}$ arising from some further physical systems with degrees of freedom completely different from spin chains and QRS, such as billiards and perturbed oscillators, and the results are included in the appendix. Again, we see robust emergence of power laws. While we do not have a systematic explanation for these features, the fact that similar patterns arise in very different systems convinces us that there is a high degree of universality underlying our empirical observations, and it certainly merits further exploration. Statistics of matrix elements of various operators in the energy eigenbasis is a central theme in relation to the Eigenstate Thermalization Hypothesis (ETH), and in particular, matrix elements in the eigenbasis of integrable Hamiltonians have been studied
in the context of departures from ETH in integrable dynamics \cite{ETHintegrable1,ETHintegrable2}. This framework is likely to be useful in exploring the origin
of universal power laws in the matrix element distributions reported here.

\section{Conclusions}

We have addressed the longstanding problem of describing quantitatively the transition regime between integrability and chaos in quantum Hamiltonian dynamics,
leading to a simple random matrix ensemble that correctly reproduces the level spacing distributions across the transition region for a variety of systems. A key ingredient of this ensemble is
the distribution of the offdiagonal matrix elements of the chaotic perturbation in the eigenbasis of the integrable part of the Hamiltonian. For these distributions,
we have uncovered remarkable, universal power-law features that invite further studies.

While we have focused here on validating our random matrix ensemble using numerical comparisons with integrability-to-chaos transition in model physical system,
an outstanding problem is to explore the features of this setup analytically. The pivotal aspects of our ensemble is the presence of a significant purely diagonal part (mimicking an integrable Hamiltonian) together with an extra random matrix term with i.i.d.\ entries. Similar ensembles are seen in the random matrix literature
\cite{RPreplica, MF}, and analytic technology is available to handle them, based on statistical field theory methods. Addressing
the question of level spacing distributions in such ensembles analytically goes beyond the current state-of-the-art, and presents an attractive problem for future work.

An explicit connection between level spacing statistics of a system whose dynamics lies between integrability and chaos and statistical properties of the chaotic
perturbation in its Hamiltonian offers an attractive novel window into quantum dynamics. Indeed, given a yet unexplored system demonstrating a crossover level
spacing statistics (i.e., neither purely Poisson nor purely WD), one will be able to extract some information about its Hamiltonian from the spectroscopic data alone. Level spacing statistics has been studied experimentally, including ultracold atomic gases \cite{Erbium} and, more recently, quantum processors \cite{processors}, with the Brody distribution employed as a fitting formula for the experimental data \cite{crossover1,crossover2}. Bringing the systematic theory we have developed here into that context will be of considerable interest.

\section*{Acknowledgments}
Work at VUB has been supported by FWO-Vlaanderen projects G012222N and G0A2226N and by the VUB Research Council through the Strategic Research Program High-Energy Physics.
MDC is supported by a Leverhulme Early Career Fellowship as well as the
Emmy Noether Fellowship program at the Perimeter Institute for Theoretical Physics and acknowledges partial support from STFC consolidated grant ST/X000664/1.
OE is supported by the  C2F program at Chulalongkorn University and by NSRF via grant number B41G680029. MP is supported by FWO-Vlaanderen through Doctoral Fellowship 1178725N.

The computational resources and services used in this work were provided by the VSC (Flemish Supercomputer Center), funded by the Research Foundation Flanders (FWO) and the Flemish Government.\vspace{1cm}

\appendix

\section{Definitions of statistical distributions}

As mentioned in the main text, the Wigner surmise captures the distribution of the level spacing statistics of random matrices that are governed by the GOE. Matrices belonging to this ensemble are real and symmetric. Their matrix elements are randomly sampled from the normal distribution $\mathcal{N}(0,1)$ for the diagonal elements, and $\mathcal{N}(0,\frac{1}{2})$ for the offdiagonal elements. An important property of this probability distribution is that it is invariant under orthogonal transformations, reflecting the independence of the choice of basis. Furthermore, the Wigner surmise is analytically derived for 2$\times$2 GOE matrices, and it matches the numerical results for large GOE matrices very well.

The RP model is a random matrix model that steps away from the GOE by removing the invariance under orthogonal transformations, through the addition of a random diagonal matrix $\mathcal{D}_{RP}$, with elements drawn from a normal distribution $\mathcal{N}(0,1)$, to a GOE matrix $\mathcal{M}_{GOE}$.
The definition of the RP model is often stated in the following form
\begin{equation}
	H = \mathcal{D}_{RP} + \frac{\nu}{D^{\gamma/2}} \mathcal{M}_{GOE} \,,
    \label{eq: RP app}
\end{equation}
with $D$ the dimension of the matrices.
In the limit $D \rightarrow \infty$, the eigenstates of the model are localized for $\gamma > 2$, featuring Poisson level statistics; for $\gamma < 1$ the matrices exhibit fully delocalized eigenstates with Wignerian statistics, while the interval $1<\gamma <2$ defines a multifractal phase, with delocalized eigenfunctions that have striking fractal structures of relevance to the phenomenon of many-body localization. In this limit, intermediate statistics occurs at $\gamma = 2$ by tuning the parameter $\nu$. Away from the large $D$ limit, $\nu$ becomes redundant and $\gamma$ can be used to dial the intensity of the GOE perturbation to induce a crossover behavior between Poisson and WD statistics. This latter parameter is then in one-to-one correspondence with $\gamma_{RP}>0$ as defined in the main text. As we pointed out in the main text, the diagonal elements of the random matrix multiplying the tunable parameter (here, $\mathcal{M}_{GOE}$) are subleading at large $D$ and have negligible impact on the level spacing statistics in the crossover region. It is therefore common (see e.g.~\cite{RPfractal}) to modify $\mathcal{M}_{GOE}$ in \eqref{eq: RP app} and set the diagonal elements of this matrix to zero. We have employed this convention in the present work since it mirrors the structure of our proposed random matrix model. 

In contrast to distributions resulting from random matrix ensembles, the Brody distribution does not originate from a theoretical framework. 
Brody's ansatz is
\begin{equation}
	P_{B,\beta}(s) = c s^\beta e^{-a s^{1+\beta}},
\end{equation}
with parameters
\begin{equation}
    a = \left[
\Gamma\!\left(\dfrac{2+\beta}{1+\beta}\right)
\right]^{1+\beta}, \qquad c = a\left(1+\beta\right)
\end{equation}
that are found by imposing unit normalization and unit average level spacing. It was proposed to incorporate fractional level repulsion and provided a good phenomenological fit to nuclear spectra.
For $\beta=0$, one recovers the Poisson distribution, while $\beta=1$ leads to the Wigner surmise.

\begin{figure*}[t]
    \centering
    \includegraphics[scale=0.6,trim= 0.5cm 0.5cm 0.5cm 0.5cm,
  clip]{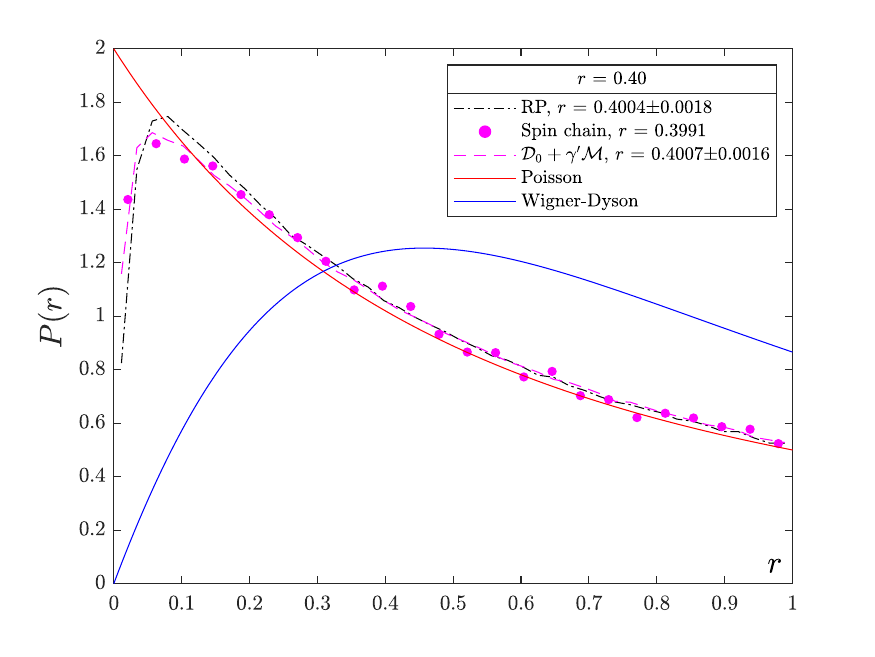} \includegraphics[scale=0.6,trim= 0.5cm 0.5cm 0.5cm 0.5cm,
  clip]{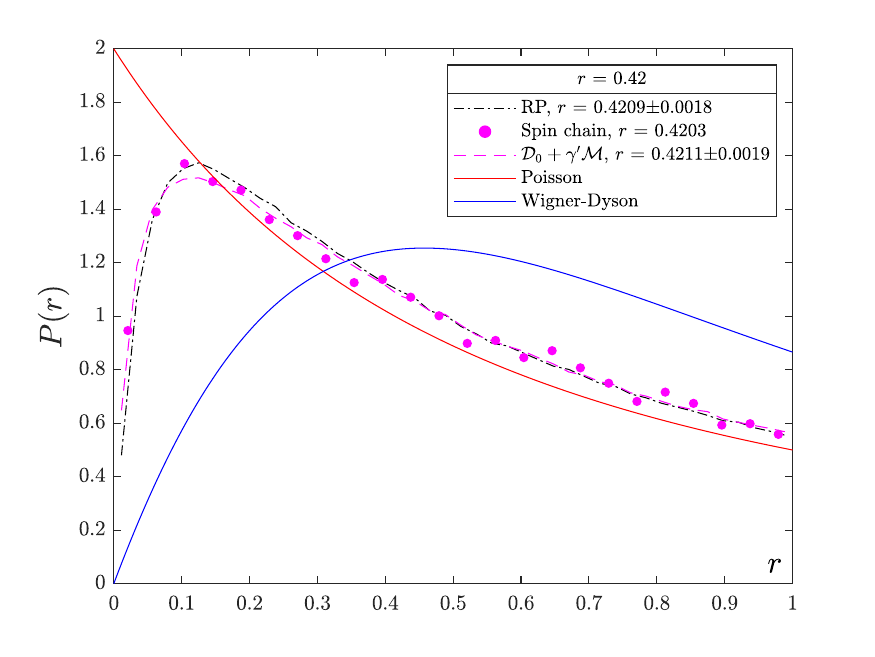}\\
  \includegraphics[scale=0.6,trim= 0.5cm 0.5cm 0.5cm 0.5cm,
  clip]{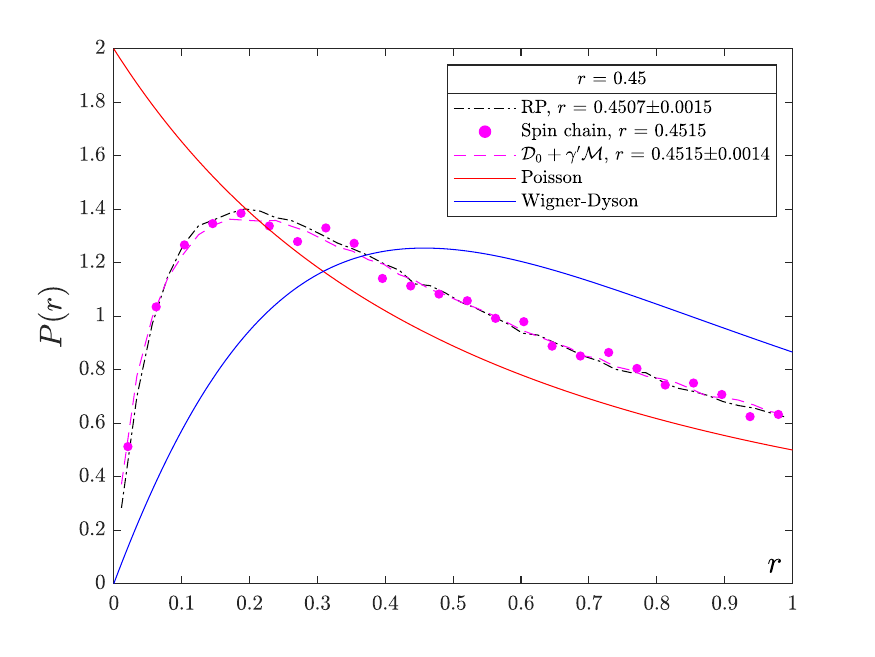} \includegraphics[scale=0.6,trim= 0.5cm 0.5cm 0.5cm 0.5cm,
  clip]{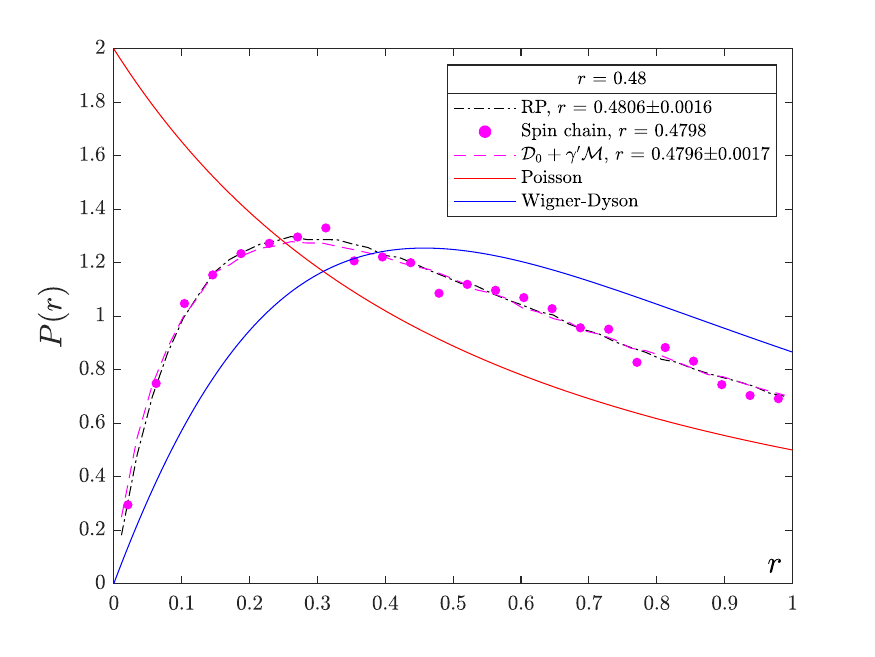}
    \caption{Distribution of $r_n$ for the spin chain model (filled circles), compared to the prediction of our random model (magenta dashed) and the RP model (black dash-dotted). Note that $r$ in the legends stands for the average value of the $r$-ratios.}
    \label{fig: Ising r-curve}
\end{figure*}

\begin{figure*}[t]
    \centering
    \includegraphics[scale=0.6,trim= 0.5cm 0.5cm 0.5cm 0.5cm,
  clip]{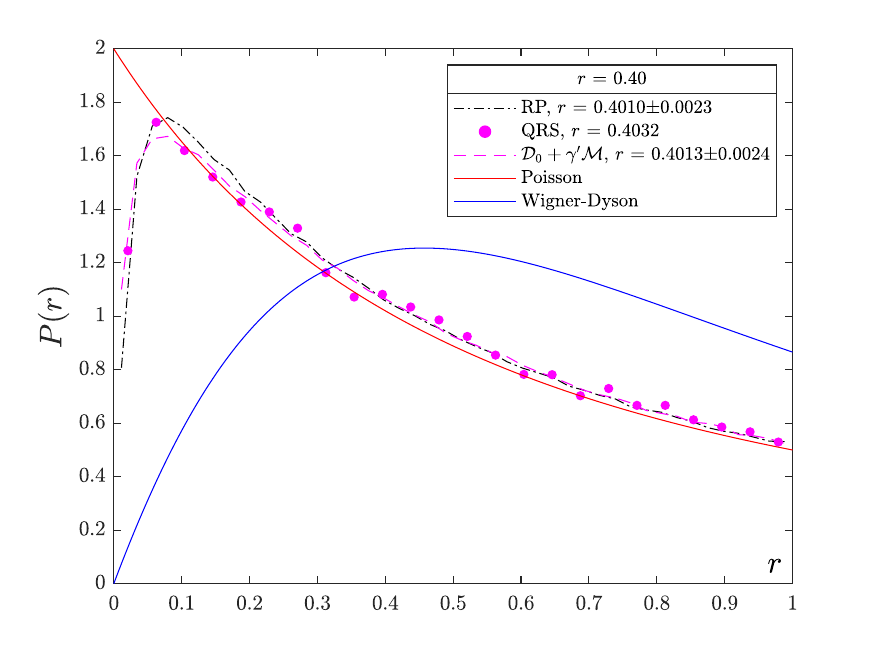} \includegraphics[scale=0.6,trim= 0.5cm 0.5cm 0.5cm 0.5cm,
  clip]{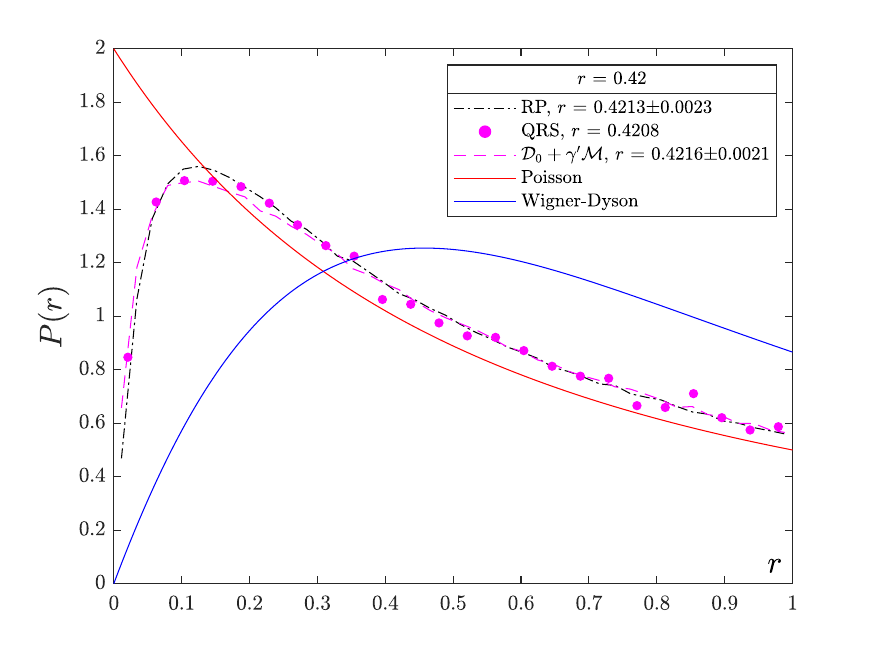}\\
  \includegraphics[scale=0.6,trim= 0.5cm 0.5cm 0.5cm 0.5cm,
  clip]{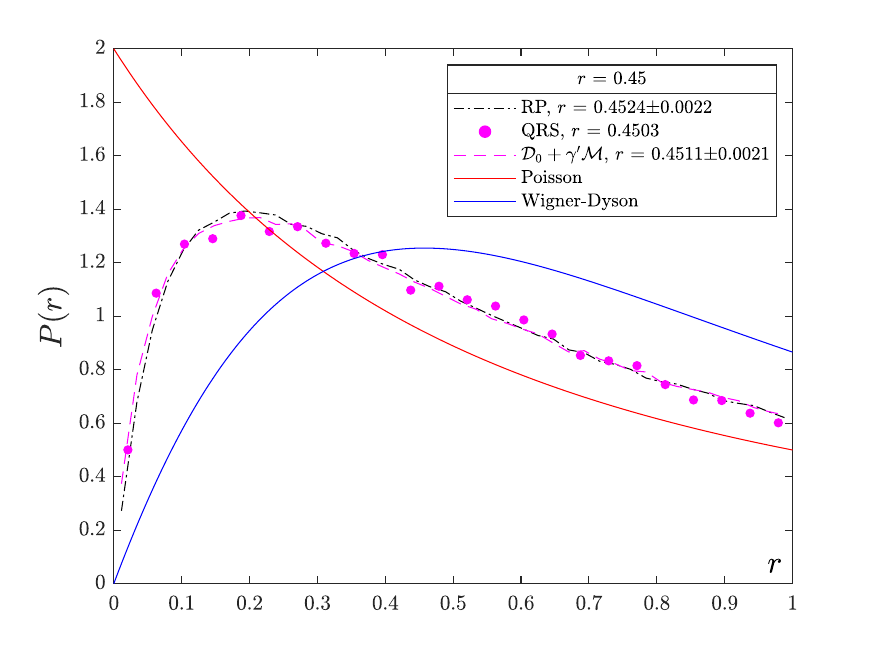} \includegraphics[scale=0.6,trim= 0.5cm 0.5cm 0.5cm 0.5cm,
  clip]{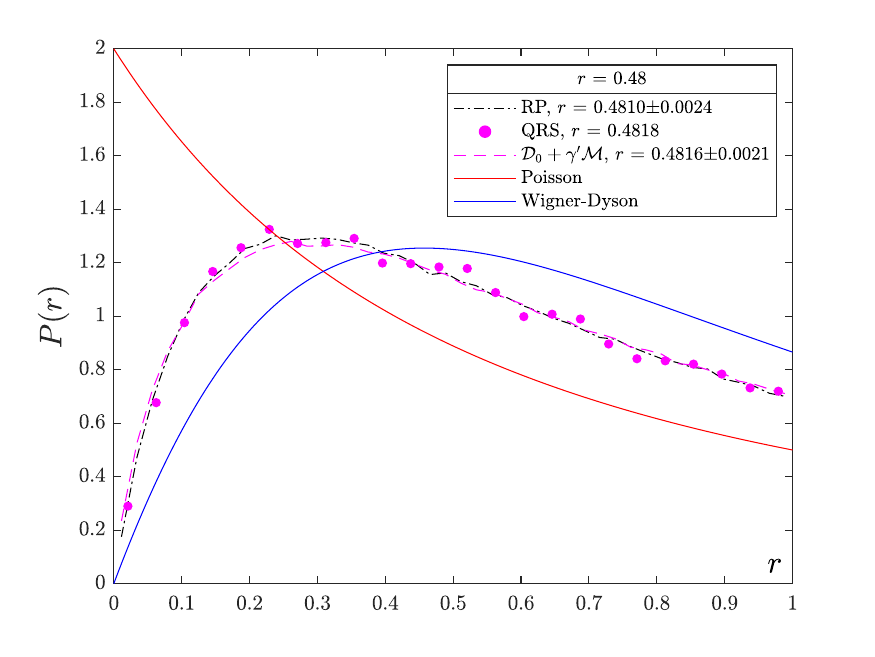}
    \caption{Distribution of $r_n$ for QRS.}
    \label{fig: QRS r-curve}
\end{figure*}

\section{Extraction of probability distributions from samples}

We will describe here the detailed construction of the probability distributions $P_{\mathcal{D}}$ and $P_{\mathcal{M}}$, the two distributions that our random matrix model defined in the main text takes as input. First, $P_{\mathcal{D}}$ is given by the empirically computed coarse grained approximation to the spectral density $\rho(E)$ of $H_0$. For both the spin chain and the QRS, this density $\rho(E)$ is close to a Gaussian, as expected \cite{nuclei}. To obtain $P_{\mathcal{M}}$, we start by writing the matrix elements of  $H_1$ in the eigenbasis of $H_0$, denoted $|n\rangle$:
\begin{equation}
	H_{1,nm} = \langle n| H_1 |m\rangle.
\end{equation}
The probability distribution $P_{\mathcal{M}}$ is then extracted from the numerical histogram obtained from the sample of all offdiagonal elements of $H_{1,nm}$. Actually, these matrix elements are very small (see Fig.~\ref{fig: powerlaw distributions}), and therefore we have extracted the numerical distribution of the logarithm (base 10) of the magnitude of the offdiagonal elements. This has proven to be a more accurate way to sample such small numbers. 

Once these histograms have been obtained from $H_0$ and $H_{1,nm}$, the random matrices $\mathcal{D}$ and $\mathcal{M}$ can be recovered by using the inverse transform sampling, which is a common method for generating random numbers from a probability distribution without an explicit formula \cite{ITS}. In order to sample variables $X$ from such distribution, one can instead use the corresponding cumulative distribution $F_X$, and sample random numbers from its inverse by using uniformly generated variables $U$ as input. These random numbers $F^{-1}_X(U)$ will have the same probability distribution as the variables $X$. We have implemented the inverse transform sampling method in MATLAB by first numerically approximating $F_X$ for the histograms of the eigenvalues of $H_0$ and offdiagonal elements of $H_{1,nm}$, respectively, using the function \texttt{cumtrapz}. In the next step, we invert the function $F_X$ by interpreting the data for $F_X$ as $x$-values and the center of the bins $X$ as $y$-values, effectively obtaining $F^{-1}_X$ by interchanging the two axes. Finally, we sample the required amount of random values $U$ from a uniform distribution between 0 and 1 and calculate the corresponding function values $F^{-1}_X(U)$ by interpolating the function $F^{-1}_X$ using the above data. For this, we use the MATLAB function \texttt{interp1}. By comparing the probability distributions of the generated random matrix elements with the original ones, we observe that the inverse transform sampling method has performed as intended.

Finally, we mention a subtlety that arises when displaying distributions on a log-scaled $x$-axis. Since we work with finite bins, these histograms naturally come with a lower cutoff. As a result, exact zeros (which would appear at $-\infty$ on a log scale) are not accounted for in our numerical procedure to construct $P_{\mathcal{D}}$ and $P_{\mathcal{M}}$. In practice, however, this is not an issue since a large majority of the (presumably) exactly zero off-diagonal elements are tiny but nonzero due to numerical errors. It is not unreasonable to assume that the elements to the left of the power-law regime in Fig.~\ref{fig: powerlaw distributions} are in fact mostly vanishing matrix elements. Only a very small fraction of the off-diagonal matrix elements (less than $0.0001\%$) are identified as exactly zero after rotating $H_1$ into the eigenbasis of $H_0$ and neglecting them does not affect our results. 

\section{$r$-ratio statistics}\label{secA1}

In recent years, an alternative to the conventional level spacing statistics has been proposed, given by  the distribution of the $r$-ratios, for which no unfolding is required, as discussed in Section 5. The analogs of the Poisson distribution and the WD distribution for the $r$-ratios are \cite{rratio1,rratio2}: 
\begin{align}
	P_P(r) &= \frac{2}{(1+r)^2}, \\
	P_W(r) &= \frac{27}{4} \frac{r+r^2}{(1+r+r^2)^{5/2}}.
\end{align}
The latter expression is a Wigner-like surmise derived from 3$\times$3 matrices of the GOE ensemble. Note that $r$ here represents the $r$-ratios, and no longer the average value of these ratios as in the main text.

As a test of our random model, we have compared the statistics of $r_n$, defined in the main text, for the two physical systems from the main text with those of our random model, as well as RP. In both cases, both RP and our model are able to match the distributions of the physical models. Our random matrix ensemble performs very well, but in this case, the numerical difference between RP and our ensemble is rather small. At the level of $r$-ratio curves, one would therefore need significantly larger system sizes to support the claim that the performance of our model is superior.

\begin{figure*}[t]
    \centering
    \includegraphics[scale=0.6,trim= 0.5cm 0.5cm 0.5cm 0.5cm,
  clip]{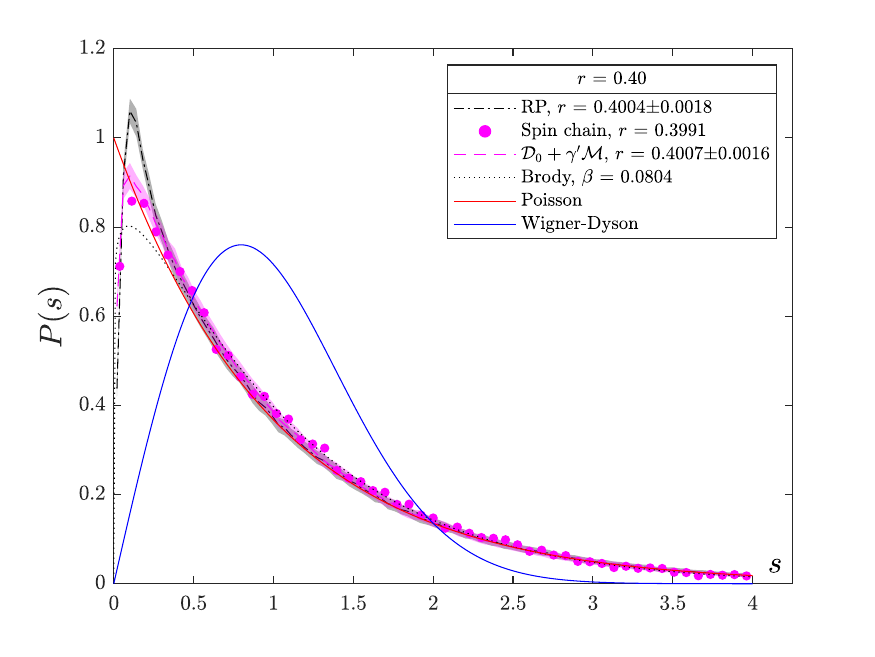} \includegraphics[scale=0.6,trim= 0.5cm 0.5cm 0.5cm 0.5cm,
  clip]{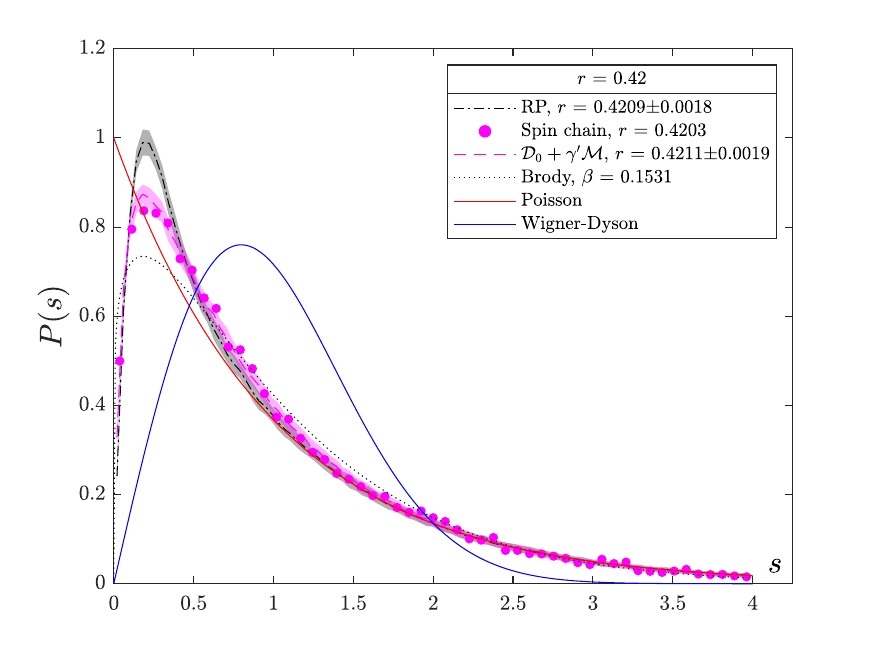}\\
  \includegraphics[scale=0.6,trim= 0.5cm 0.5cm 0.5cm 0.5cm,
  clip]{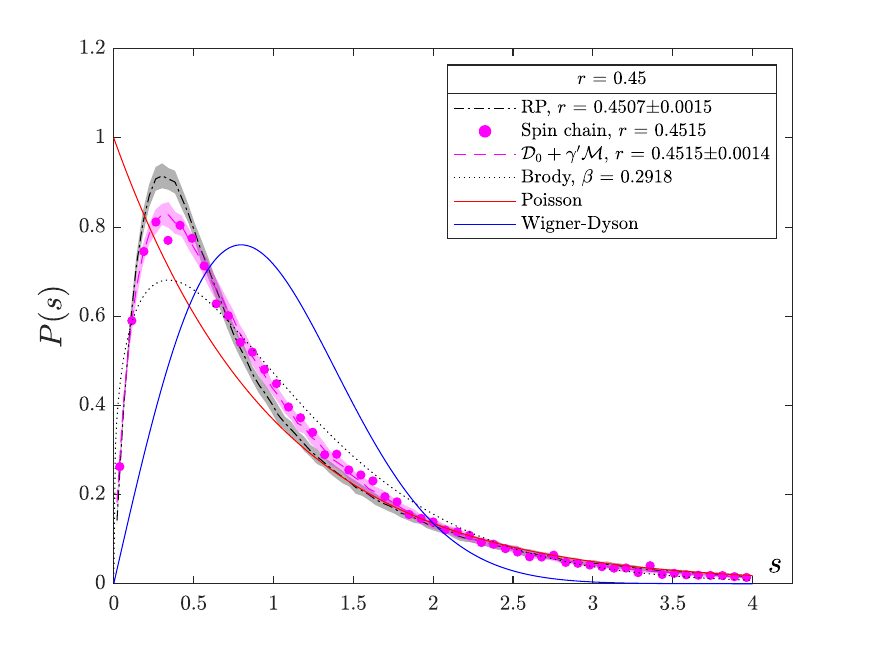} \includegraphics[scale=0.6,trim= 0.5cm 0.5cm 0.5cm 0.5cm,
  clip]{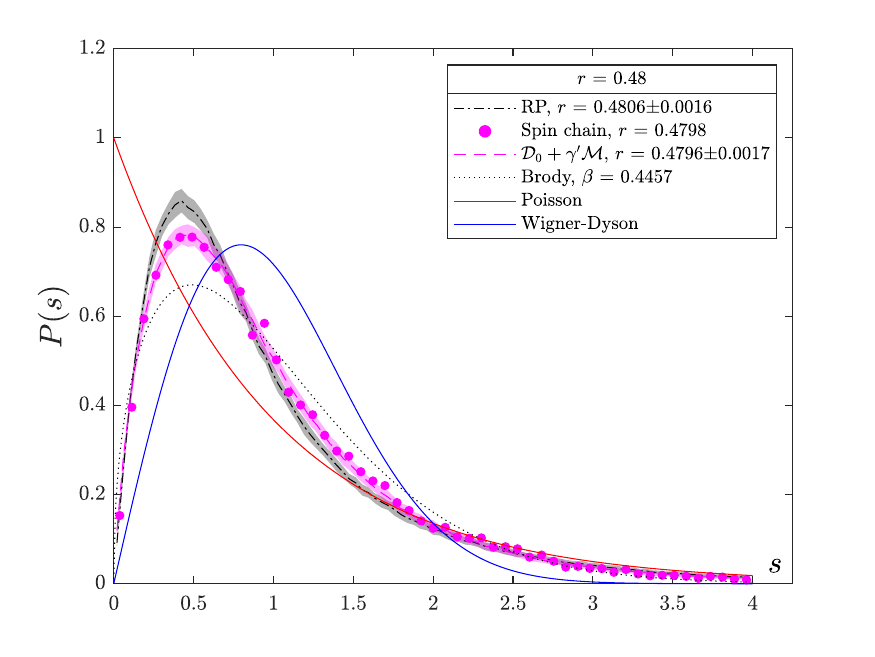}
    \caption{The same data as in Fig.~\ref{fig: Ising histograms}, but now with error bars included for the random models, depicted as a belt of width $2\sigma$.}
    \label{fig: Ising histograms+errorbar}
\end{figure*}
\begin{figure*}[t]
    \centering
    \includegraphics[scale=0.6,trim= 0.5cm 0.5cm 0.5cm 0.5cm,
  clip]{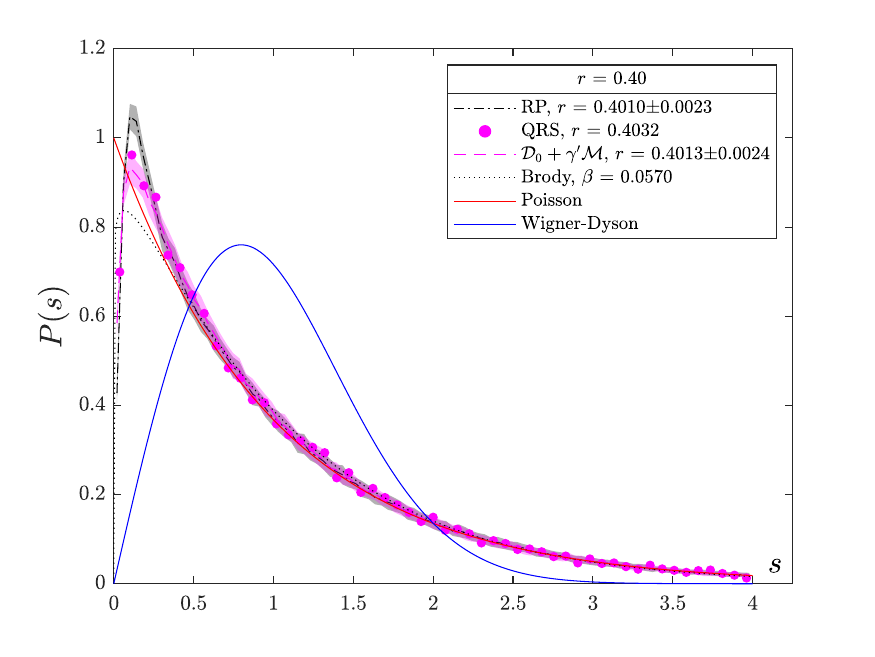} \includegraphics[scale=0.6,trim= 0.5cm 0.5cm 0.5cm 0.5cm,
  clip]{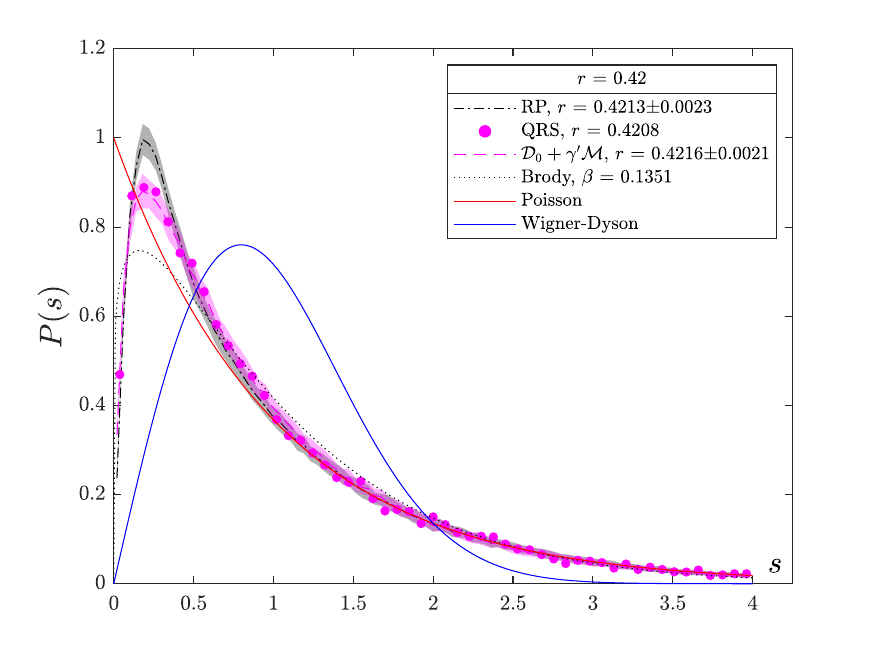}\\
  \includegraphics[scale=0.6,trim= 0.5cm 0.5cm 0.5cm 0.5cm,
  clip]{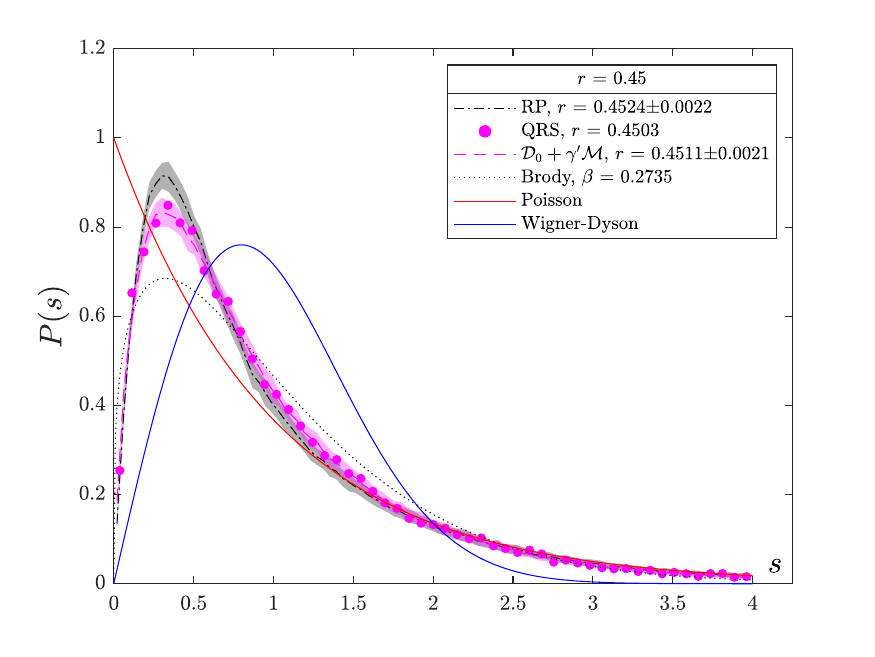} \includegraphics[scale=0.6,trim= 0.5cm 0.5cm 0.5cm 0.5cm,
  clip]{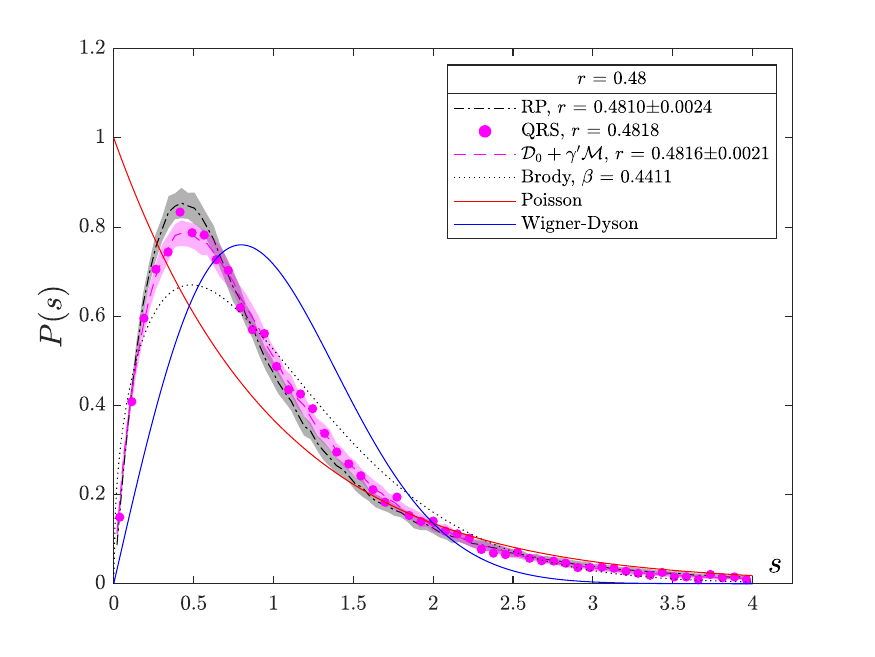}
    \caption{The same data as in Fig.~\ref{fig: QRS histograms}, but now with error bars included for the random models, depicted as a belt of width $2\sigma$.}
    \label{fig: QRS histograms+errorbar}
\end{figure*}

\section{Error bars on level spacings from random matrix ensembles}

At finite $D$, the level spacing statistics of a single realization of a random matrix model depends on the draw. In order to obtain stable statistics to compare with the level spacing statistics of the physical Hamiltonian, it is common practice to lump together the spacings from many realizations of the ensemble, thereby obtaining an average distribution (as was done in the main text). It is reasonable to expect the fluctuations around the average curve to decrease with increasing matrix dimension. Here, we provide more information about the variance on these averaged distributions. Figs.~\ref{fig: Ising histograms+errorbar} and \ref{fig: QRS histograms+errorbar} display the same data as in the main text, with the addition of a belt of width $2\sigma$ surrounding the spectra of our random model ($1\sigma$ above and below the average) and the RP model. The first point to note is the persistent observable difference between the RP model and our random model at most values of $s$ and $r$. Second, most of the data points for the physical systems lie within a standard deviation from the averaged curve produced by our random matrix model, showcasing the success of the approach. We verified numerically that the width of the belt decreases as $D$ increases in both random models, as expected.

\section{Power laws in matrix element distributions}

We have shown in the main text that the distributions of the offdiagonal elements of $H_1$ for both the Ising chain and the QRS model display a power law regime that extends over multiple orders of magnitude. The appearance of this behavior in both models has led us to check if it is also present in other types of systems. We show the results here for a billiard system and a perturbed oscillator.

\begin{figure*}[t]
\centering
\includegraphics[scale=0.4]{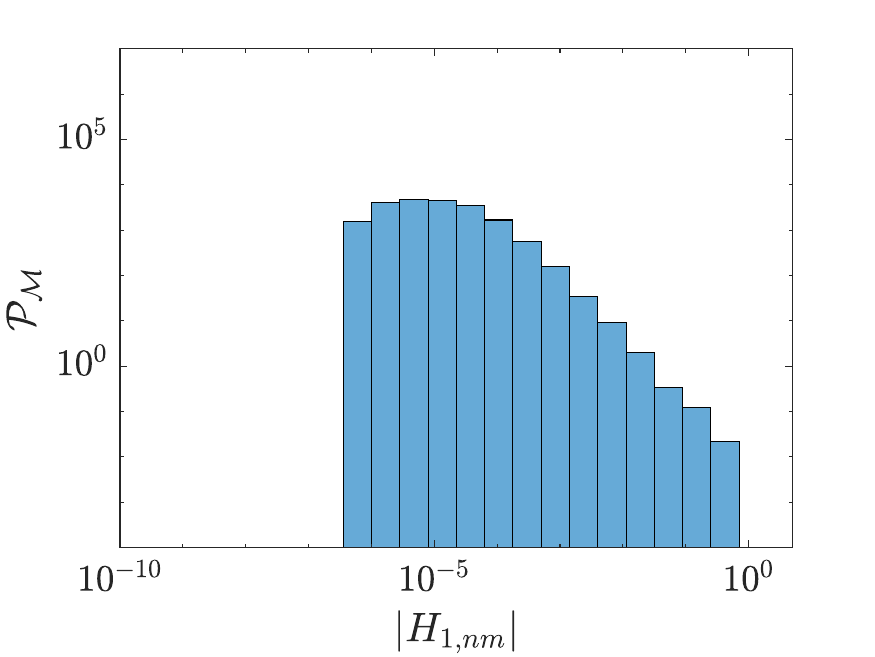}\includegraphics[scale=0.4]{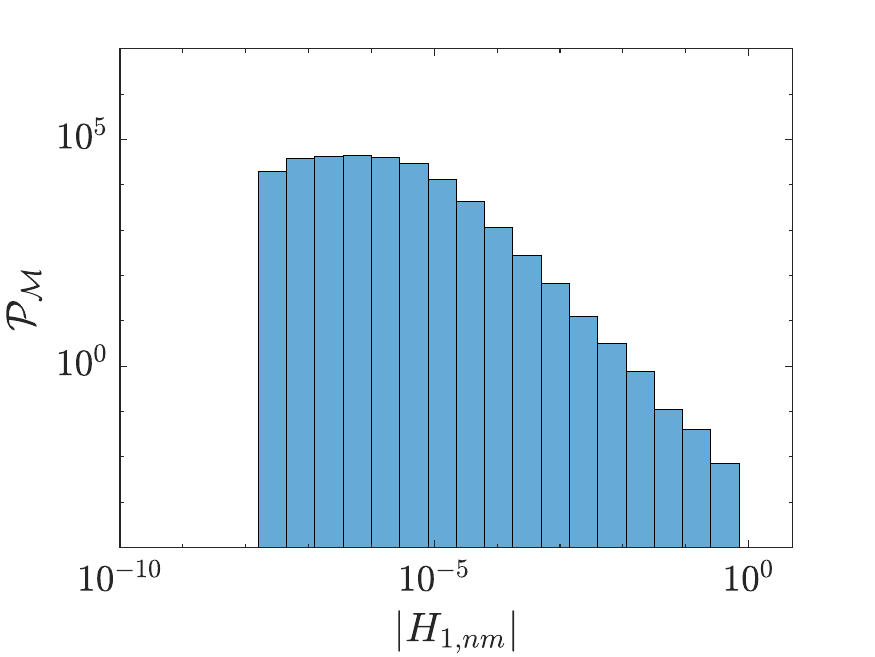}\includegraphics[scale=0.4]{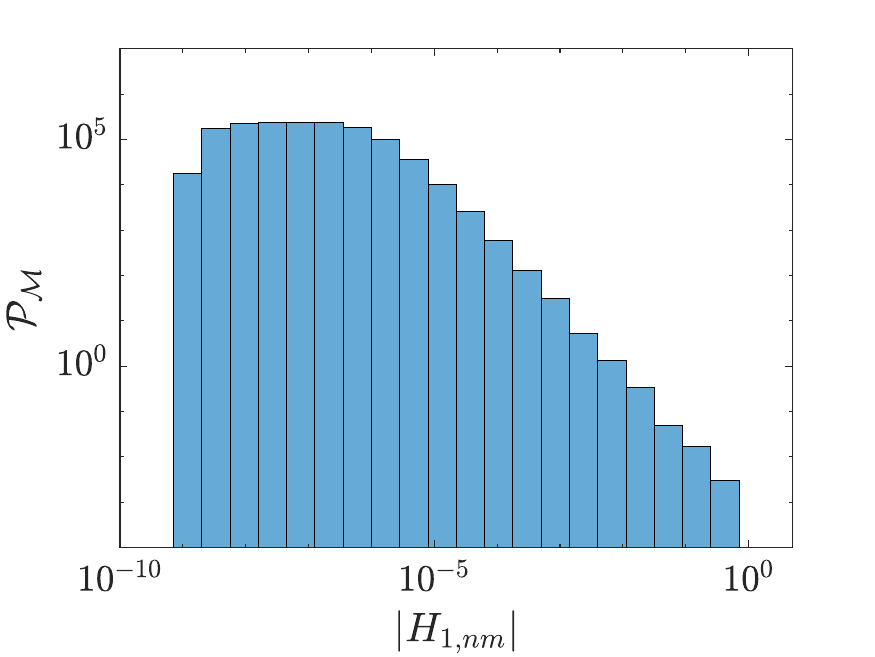} 
\caption{Probability distribution of the offdiagonal matrix elements of $H_1$ in the eigenbasis of $H_0$ in \eqref{eq: Hbilliard}. Left: $E_{\max}=100^2 \pi^2$ ($\sim 10^6$ nonzero matrix elements), Middle: $E_{\max}=300^2 \pi^2$  ($\sim 3 \times 10^7$ nonzero matrix elements), Right: $E_{\max}=700^2 \pi^2$ ($\sim 4 \times 10^8$ nonzero matrix elements). A power law regime is clearly visible, and it extends when taking more eigenstates into account. }
\label{fig: distributions billiards}
\end{figure*}

First, we consider a rectangular billiard with length $L_x=\sqrt2$ in the $x$-direction and $L_y=1$ in the $y$-direction, with a harmonic potential perturbation:
\begin{equation}
\label{eq: Hbilliard}
    H = \underbrace{p_x^2+p_y^2}_{\text{$H_0$}} + \underbrace{x^2+y^2}_{\text{$H_1$}} \, .
\end{equation}
The eigenstates of the integrable model $H_0$ are labeled by two quantum numbers $(n,m)$ with eigenstates
\begin{equation}
    \psi_{nm}(x,y) = \sqrt{2\sqrt{2}}\sin \frac{ \pi n x}{\sqrt{2}}\sin \pi m y \, ,
\end{equation}
and energies $E_{nm} = \left(\frac{n^2}{2} + m^2 \right) \pi^2$.
We consider the distributions of offdiagonal elements of the harmonic potential in these integrable eigenstates, for different pairs $(n,m)$ and $(n',m')$
\begin{align}
    \int & dx dy  \, \psi_{nm}(x,y) \left( x^2+y^2 \right) \psi_{n'm'}(x,y) \nonumber \\
    & = a_{nn'}(\sqrt{2}) \delta_{mm'}+a_{mm'}(1) \delta_{nn'}
    \label{eq: offdiag billiard}
\end{align}
with $a_{nn'}(L) = \frac{8 (-1)^{n + n'} n n' L^2 }{(n - n')^2 (n + n')^2 \pi^2}$, where we assumed $n\neq n'$.
Note that many offdiagonal elements are zero, so we will focus on the nonzero elements only (and take the absolute value), which are shown in Fig.~\ref{fig: distributions billiards}. 
\begin{figure*}[t]
\centering
\includegraphics[scale=0.4]{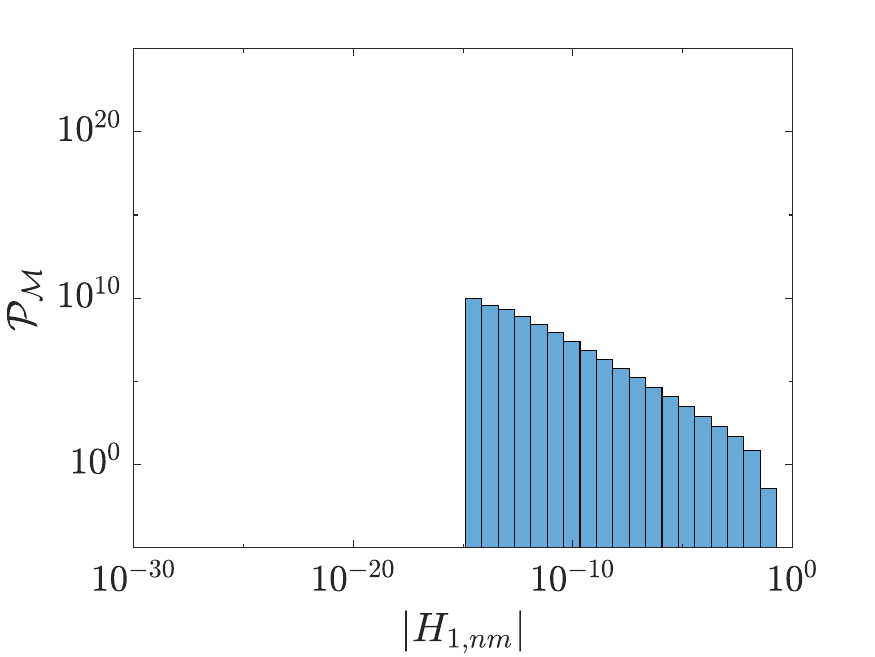}\includegraphics[scale=0.4]{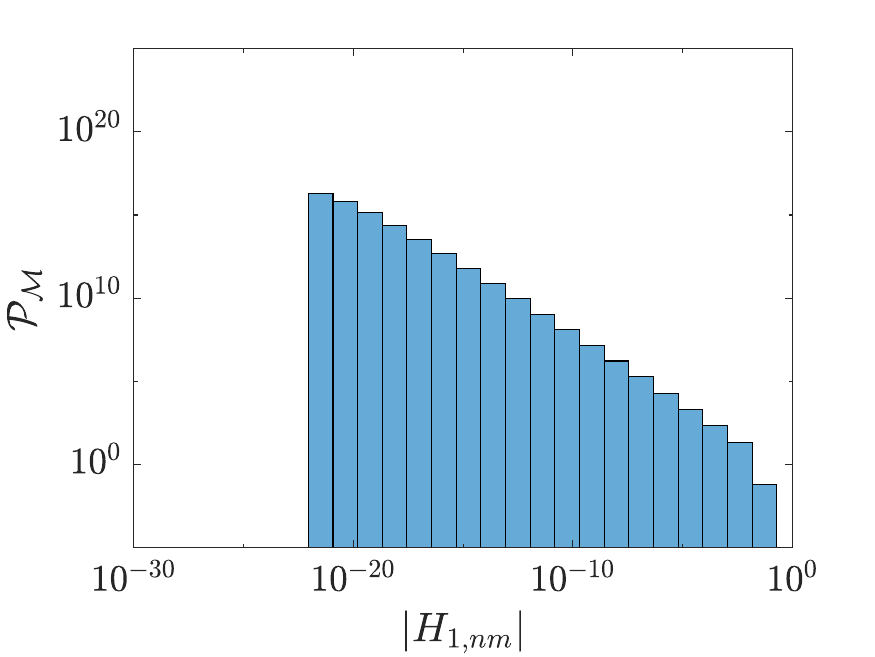}\includegraphics[scale=0.4]{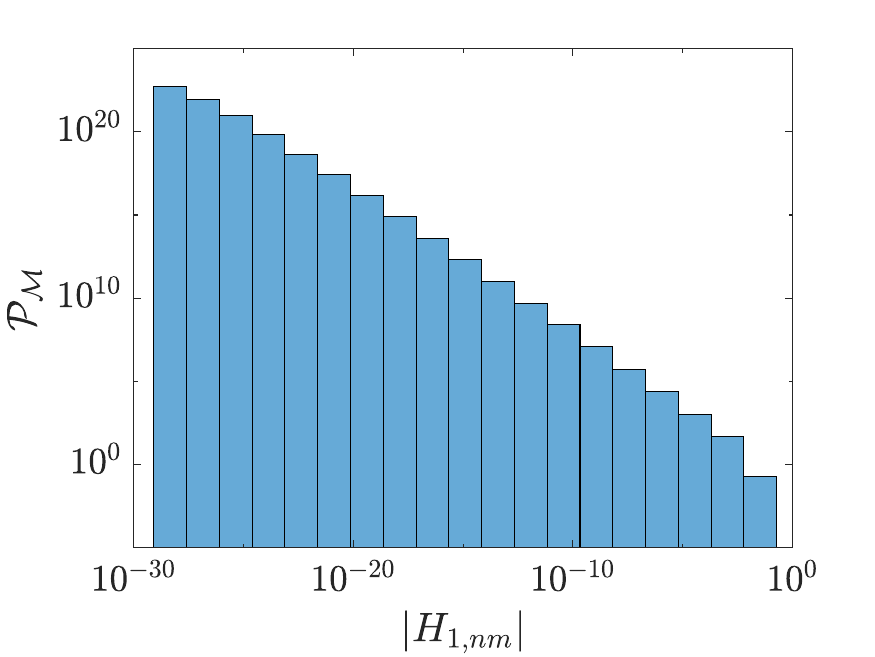}  
\caption{Probability distribution of the offdiagonal matrix elements of the perturbation in the eigenbasis of the integrable model for the anisotropic oscillator in \eqref{eq: osc}. Left: $E_{\max}=100$ (97143 nonzero matrix elements), Middle: $E_{\max}=150$  (493289 nonzero matrix elements), Right: $E_{\max}=200$ (1560714 nonzero matrix elements). The power law regime appears for all three examples, and extends over a wider domain as $E_{\max}$ increases.}
\label{fig: distributions osc}
\end{figure*}
There is an initial increase at low values, followed by a peak and a power law decrease. As we take more eigenstates into account by increasing the energy cutoff, the peak moves to the left and the power law regime appears to extend. 

Given the simplicity of the expression $a_{nn'}(L)$, an analytic understanding of the shape of these distributions at infinite cutoff can be obtained as follows. Focus for conciseness on the offdiagonal elements for which $n=n'$, varying $m$ and $m'<m$. (The case $m=m'$ is similar.) One starts by defining the variables $u=m \pi/\sqrt{E_{\max}}$, $v=m' \pi/\sqrt{E_{\max}}$ and $z=n  \pi/\sqrt{2E_{\max}}$ that become continuous in the limit $E_{\max} \rightarrow \infty$ and lie within the interval $[0,1]$. The matrix elements now take the form of a continuous function $a_{mm'} \rightarrow a(u,v)$. Then, the probability distribution $P(x)$ for offdiagonal elements taking value between $x$ and $x+dx$ can be approximated by 
\begin{equation}
    P(x) \sim \int_0^{1} dz \int_0^{\sqrt{1-z^2}}\hspace{-2.8em} du \hspace{1.2em} \int_{0}^{u}  \delta(a(u,v)-x)dv \,.
\end{equation}
Working out these integrals, one finds the scalings
\begin{equation}
          P(x)\sim \begin{cases}
       \text{constant }& \text{for }x \lesssim 1/E_{\max}, \\
     x^{-3/2}       & \text{for } x \gtrsim 1/E_{\max}.
    \end{cases}
    \end{equation}
The slopes of the histograms seen in Figs.~\ref{fig: distributions billiards} can be verified to converge slowly to the analytically derived exponent of the power law regime. Note that the power law exponent (understood as $p$ in $x^{-p}$) is outside the range $0<p<1$ that was relevant for our analysis of the many-body systems in the main text.

As a second example, we perform a similar numerical analysis for a perturbed anisotropic harmonic oscillator 
\begin{equation}
\label{eq: osc}
	H = \underbrace{p_x^2 + p_y^2 + x^2 + 2y^2}_{\text{$H_0$}} + \underbrace{e^{-x^2-y^2}}_{\text{$H_1$}}.
\end{equation}
The eigenfunctions $\psi_{nm}(x,y)$ of $H_0$ are
\begin{equation}
	\psi_{nm}(x,y) = \frac{\sqrt[8]{2}\,e^{\frac{-x^2-\sqrt{2}y^2}{2}}}{\sqrt{\pi2^n 2^m n! m!}} h_n(x)h_m(\sqrt[4]{2}y),
\end{equation}
where $h_n(x)$ represents the Hermite polynomials, with energies $E_{nm}=2n+1+\sqrt{2}(2m+1)$. 
An analytic expression for the matrix elements of $H_1$ in the eigenbasis of $H_0$ can be obtained by evaluating
\begin{align}
\label{eq: offdiags osc}
	\int & dx \int dy ~\psi_{nm}(x,y)~ e^{-x^2-y^2} ~\psi_{n'm'}(x,y) \\
    &= f_{nn'}(1)f_{mm'}(\sqrt{2})\nonumber
\end{align}
with
\begin{equation}
    f_{nn'}(a) = \begin{cases}
    \frac{\sqrt{n!n'!a}(-1)^n}{(-2)^{\frac{n+n'}{2}}} s_{nn'}(a)& \text{for }n+n' \text{ even}, \\
    0              & \text{otherwise},
    \end{cases}
\end{equation}
and 
\begin{equation}
    s_{nn'}(a)=\left( 1+a\right)^{-\frac{n+n'+1}{2}}\sum_{k} \frac{(2a)^k}{k!\left(\frac{n-k}{2}\right)!\left(\frac{n'-k}{2}\right)!}\, ,
\end{equation}
with $k$ running between $0,2,\dots,\min(n,n')$ if $n$ and $n'$ are even or $1,3,\dots,\min(n,n')$ if $n$ and $n'$ are odd.
Fig.~\ref{fig: distributions osc} shows the histogram for the absolute value of the offdiagonal elements of $H_{1,nm}$ with an upper bound $E_{\max} = 100,150,200$ on the energies $E_{nm}$, clearly displaying a power law regime.

\end{document}